\documentclass[12pt,epsf,qsymbols]{article}
\pdfoutput=1
\usepackage[utf8]{inputenc}
\usepackage[normalem]{ulem}
\usepackage[english]{babel}
\usepackage{tabularx}
\usepackage{array}
\usepackage{soul}
\usepackage{graphics}
\usepackage[pdftex]{graphicx}
\usepackage{psfrag}
\usepackage{epsfig}
\usepackage{subfig}
\usepackage{mathtools}
\usepackage{amssymb}
\usepackage{bbold}
\usepackage{setspace}
\usepackage{rotating}
\usepackage{colortbl}
\usepackage{longtable}
\usepackage{slashed}
\usepackage{braket}
\usepackage{lineno}
\makeatletter
\usepackage{textcomp}
\usepackage[usenames,dvipsnames]{xcolor}
\usepackage{relsize}
\usepackage{cite}

\usepackage{float}

\newcommand{\specialcell}[2][c]{%
\begin{tabular}[#1]{@{}c@{}}#2\end{tabular}}

\usepackage{bbm}
\usepackage{bm}
\usepackage{enumerate}
\usepackage{amsmath}
\usepackage{verbatim}

\usepackage{hyperref}
\hypersetup{colorlinks,citecolor=nicegreen,linkcolor=niceblue}
\hypersetup{colorlinks=true}

\usepackage{pifont}

\usepackage{makecell}

\usepackage{arydshln}

\allowdisplaybreaks

\setlongtables

\newcommand{\bea}{\begin{eqnarray}}
\newcommand{\eea}{\end{eqnarray}}
\newcommand{\beq}{\begin{equation}}
{
\newcommand{\eeq}{\end{equation}}
\newcommand{\ec}{\end{center}}
\newcommand{\bc}{\begin{center}}

\newcommand{\pdir}{p\kern -5.2pt\raise 0.2ex\hbox {/}}

\newcommand{\vdir}{v\kern -5.75pt\raise 0.15ex\hbox {/}}
\newcommand{\kdir}{k\kern -5.75pt\raise 0.15ex\hbox {/}}
\newcommand{\epsdir}{\epsilon\kern -5.0pt\raise 0.15ex\hbox {/}}
\newcommand{\bvdir}{\bar{v}\kern -5.75pt\raise 0.15ex\hbox {/}}
\newcommand{\Ddir}{D\kern -7.75pt\raise 0.20ex\hbox {/}}
\newcommand{\Adir}{A\kern -7.75pt\raise 0.20ex\hbox {/}}
\newcommand{\ldir}{l\kern -5.0pt\raise 0.2ex\hbox{/}}
\newcommand{\varepsdir}{\varepsilon\kern -5.5pt\raise 0.15ex\hbox{/}}

\makeatother

\definecolor{niceblue}{rgb}{0.15,0.15,0.6}
\definecolor{nicegreen}{rgb}{0.1,0.5,0.1}
\definecolor{Red}{rgb}{1.,0.,0.}

\definecolor{Green}{rgb}{0.2,.7,0.2}

\begin{document}
\unitlength = 1mm

\thispagestyle{empty} 
\begin{flushright}
\begin{tabular}{l}
\end{tabular}
\end{flushright}
\begin{center}
\vskip 3.4cm\par
{\par\centering \textbf{\LARGE  
\Large \bf Implications of scalar and tensor explanations of $R_{D^{(\ast)}}$}}
\vskip 1.2cm\par
{\scalebox{.85}{\par\centering \large  
\sc Ferruccio Feruglio$^{a,b}$, Paride Paradisi$^{a,b}$ and Olcyr Sumensari$^{a,b}$}
{\par\centering \vskip 0.7 cm\par}
{\sl 
$^a$~Istituto Nazionale Fisica Nucleare, Sezione di Padova, I-35131 Padova, Italy}\\
{\par\centering \vskip 0.2 cm\par}
{\sl 
$^b$~Dipartamento di Fisica e Astronomia ``G.~Galilei", Università di Padova, Italy}\\

{\vskip 1.65cm\par}}
\end{center}

\vskip 0.85cm
\begin{abstract}
We investigate the implications of scalar and tensor operators proposed to accommodate the hints of lepton flavor universality violation in charged-current $B$-meson decays. We show that these scenarios unavoidably induce chirality enhanced contributions to charged lepton magnetic moments and Yukawa couplings. By using an effective field theory approach we quantify in a model independent way the connection of the $R_{D^{(\ast)}}$ anomaly with the tau $g-2$ and the Higgs decay $h\to \tau\tau$, which can offer an alternative to test these scenarios. Concrete New Physics models giving rise to our setup will be illustrated.
\end{abstract}
\newpage
\setcounter{page}{1}
\setcounter{footnote}{0}
\setcounter{equation}{0}
\noindent

\renewcommand{\thefootnote}{\arabic{footnote}}

\setcounter{footnote}{0}


\newpage

\section{Introduction}
\label{sec:intro}
Data accumulated in the last years show an excess of $B\to D^{(*)} \tau \nu$ decays over the Standard Model (SM) predictions, hinting at a violation of lepton flavor universality (LFU) in charged-current $b\to c$ transitions~\cite{Lees:2012xj,Lees:2013uzd,Huschle:2015rga,Aaij:2015yra,Hirose:2016wfn,Sato:2016svk,Abdesselam:2016cgx}. While no single measurements is conclusive yet, all central values are coherently shifted above the SM ones and global fits indicate a tension at the 4\,$\sigma$ level~\cite{Amhis:2016xyh}. If this discrepancy is confirmed by future data, it will represent a major discovery, contradicting many of our prejudices which suggest New Physics (NP) hidden in loop and CKM-suppressed transitions and not in tree-level CKM-favored processes. Hints of LFU violations are also present in neutral current $b\to s$ transitions,
where a deficit of muons seems to occur in semileptonic exclusive decays of the $B$ meson into $K/K^*$~\cite{Aaij:2014ora,Aaij:2017vbb}. This data have prompted a large number of theoretical interpretations with tantalizing simultaneous explanations of both anomalies~\cite{Bhattacharya:2014wla,Glashow:2014iga,Greljo:2015mma,Alonso:2015sja,Fajfer:2015ycq,Falkowski:2015zwa,Barbieri:2015yvd,Hiller:2016kry,Barbieri:2016las,Assad:2017iib,DiLuzio:2017vat,Calibbi:2017qbu,Bordone:2017bld,DAmbrosio:2017wis,Barbieri:2017tuq,Becirevic:2016yqi,Becirevic:2018afm,Marzocca:2018wcf,Blanke:2018sro,Altmannshofer:2017poe}. The scale $\Lambda$ of NP implied by the data is relatively small~\cite{DiLuzio:2017chi}. In particular, for the charged-current anomaly, $\Lambda$ is required to be close to the TeV scale, exposing all the existing interpretations to a large variety of constraints, ranging from collider bounds coming from the overproduction of high-$p_T$ lepton pairs~\cite{Faroughy:2016osc}, to bounds related to LFU tests in pure leptonic processes~\cite{Feruglio:2016gvd}.

These last constraints can be accounted for by analysing the renormalization group evolutions (RGE) of the semileptonic operators invoked at the scale $\Lambda$ to explain the anomalies. Such an analysis has been already carried out for operators of the type (current)$\times$(current) in Ref.~\cite{Feruglio:2016gvd,Feruglio:2017rjo,Cornella:2018tfd}. In this case the running is dominated by electroweak effects and, in the simplest case of NP mainly affecting the third fermion generation, a simultaneous explanation of all $B$-physics anomalies is severely challenged by the existing bounds. In a full-fledged model RGE outcomes represent an important contribution, which however should be combined with other UV effects, leaving open the possibility of partial cancellations.
Moreover they can be suppressed or made inoffensive by raising the scale $\Lambda$ and/or by adopting more general flavor patterns, where
the couplings to lighter generations are not necessarily dominated by the mixing with the third one. In general, even in this more general context, radiative
effects have been proven important to assess the validity of a given model and an RGE analysis based on an effective field theory (EFT) approach, though not complete, is certainly very useful as a first guide.

In this paper, we will perform a similar analysis focusing on the scalar/tensor semileptonic operators, which can accommodate the anomalies in charged currents. This will nicely complement the existing discussion for the (current)$\times$(current) operators. Our study is motivated by the fact that popular models addressing the $B$-anomalies include leptoquarks (LQ) in their spectrum, whose exchange can give rise to the operators analyzed here~\cite{Dorsner:2016wpm}. Moreover, some scalar LQs are known to be good candidates for an explanation of the muon $g-2$ anomaly with LQ masses in the TeV range and coupling constants in a weak coupling regime~\cite{Biggio:2014ela,Biggio:2016wyy}. In order to achieve quantitative results, we need to formulate some assumptions, which define our benchmark scenario.
We will assume a reference flavor pattern where NP only couples to the third quark generation in the interaction basis, 
turning on the minimum amount of mixing needed to feed the effect to the second quark generation.
In the lepton sector we allow diagonal NP couplings to all generations, motivated by the possibility of accommodating the discrepancy in the muon $g-2$. We will separately comment on the impact of non-diagonal lepton couplings. 

The main outcome of our study is that a chirally enhanced contribution to the charged lepton magnetic moments is an unavoidable consequence of the RGE flow starting from tensor operators. Moreover, the presence of scalar operators involving leptons of the third generation gives rise to modified couplings of the Higgs boson to $\tau$'s, at a level which is not far from the present experimental accuracy. Scalar and tensor operators in relation to $B$-anomalies have already been investigated in Ref.~\cite{Becirevic:2012jf,Biancofiore:2013ki,Freytsis:2015qca,Becirevic:2016hea,Celis:2016azn,Bardhan:2016uhr,Ivanov:2017mrj}. The former are known to be constrained by the $B_c$ lifetime through the enhancement of the $B_c^-\to \tau^- \bar{\nu}$ channel \cite{Alonso:2016oyd}. Renormalization of scalar and tensor operators, their strong mixing and the impact on phenomenology has been recently analyzed in Ref.~\cite{Gonzalez-Alonso:2017iyc}, where charged lepton magnetic moments and Higgs decays properties, which are the main focus of the present analysis, have not been addressed. Furthermore, contributions to the lepton magnetic moments in specific LQ models have been studied in Ref.~\cite{Biggio:2014ela,Bauer:2015knc,Cai:2017wry,ColuccioLeskow:2016dox}. Here, we stress their tight relation to tensor operators, whatever UV origin they may have.

\
 
The remainder of this paper is organized as follows. In Section~\ref{ssec:eff-lagrangian} we define our framework and we discuss the RGE effects induced by the running from the heavy scale $\Lambda$ down to lower energy scales. In Section~\ref{sec:pheno} we discuss individually the different physical quantities affected in our framework. In Section~\ref{sec:numerical} we analyze globally the available parameter space and quantify our numerical predictions. In Section~\ref{sec:NP_models} we briefly comment on the concrete NP scenarios giving rise to the operators we consider. Finally, in Section~\ref{sec:conclusion} we will summarize and comment our results.

	\section{Effective Lagrangian}
	\label{ssec:eff-lagrangian}

Our starting point is an effective Lagrangian $\mathcal{L}$ defined at the scale $\Lambda \approx 1$~TeV extending the SM one 
by the addition of a restricted set of dimension six semileptonic operators:
\begin{equation}
\label{Lagr}
\mathcal{L}=\mathcal{L}_{\mathrm{SM}}+\mathcal{L}_{\mathrm{NP}}^0\,,
\end{equation}
where the NP contribution is given by
%

\begin{equation}
\label{eq:LNP0}
\mathcal{L}_{\mathrm{NP}}^0 = \dfrac{C_{S_R}^{prst}}{\Lambda^2}\, \big{[}\mathcal{O}_{\ell e qd}\big{]}_{prst}+\dfrac{C_{S_L}^{prst}}{\Lambda^2}\, \big{[}\mathcal{O}^{(1)}_{\ell e qu}\big{]}_{prst}+\dfrac{C_{T}^{prst}}{\Lambda^2}\, \big{[}\mathcal{O}^{(3)}_{\ell e qu}\big{]}_{prst}+\mathrm{h.c.}\,,
\end{equation}

\noindent where $p,r,s,t$ are flavor indices, and $C_i$ ($i\in \lbrace S_L, S_R, T \rbrace$) are generic Wilson coefficients of the operators~\cite{Buchmuller:1985jz,Grzadkowski:2010es}
\begin{align}
\begin{split}
\big{[}\mathcal{O}_{\ell e qd}\big{]}_{prst} &=  \big{(}\overline{L^{\prime}_p}^a e^\prime_{rR}\big{)} \big{(}\overline{d^\prime_{sR}}  {Q^\prime_t}^a\big{)}\,,\\[0.3em]
\big{[}\mathcal{O}_{\ell e qu}^{(1)}\big{]}_{prst} &= \big{(}\overline{L^\prime_p}^a e^{\prime}_{rR}\big{)} \varepsilon_{ab} \big{(}\overline{Q^\prime_s}^b  u^\prime_{tR}\big{)}\,,\\[0.3em]
\big{[}\mathcal{O}_{\ell e qu}^{(3)}\big{]}_{prst} &= \big{(}\overline{L^\prime_p}^a \sigma_{\mu\nu}e^\prime_{rR}\big{)} \varepsilon_{ab} \big{(}\overline{Q^\prime_s}^b \sigma^{\mu\nu} u^\prime_{Rt}\big{)}\,.
\end{split}
\end{align} 

\noindent Here $a,b$ are $SU(2)_L$ indices, and $\varepsilon_{12}=-\varepsilon_{21}=1$. Primed fields refer to the interaction basis. We do not consider semileptonic operators of the type (current)$\times$(current) since their low-energy effects have been already studied. Throughout this paper we assume that NP only affects the third quark generation, while for leptons we will be less restrictive. Indeed, as far as quarks are concerned, this scenario is favored by model building, in particular in the framework of flavor symmetries. Moreover, as we will see, NP couplings to the third quark generation is sufficient to explain the charged-current anomalies without conflicting with other existing data. On the other hand, we allow NP couplings to the first and second lepton generations to investigate the implications on the lepton magnetic moments. In particular, we consider the flavor structure

\begin{equation}
C_{i}^{prst} = \delta_{3s}\, \delta_{3t}  \,\, C_i^{pr}\,,
\end{equation} 

\noindent where the matrix $C_i^{pr}$ specifies the flavor pattern in the lepton sector. We will consider the case of diagonal entries for $C_i^{pr}$, namely,

\begin{equation}
C_i = \begin{pmatrix}
C_i^e & 0 & 0\\ 
0 & C_i^\mu & 0\\ 
0 & 0 & C_i^\tau
\end{pmatrix}\,,
\end{equation}

\noindent where $C_i^k$ ($k=e,\mu,\tau$) are arbitrary effective coefficients, which we assume to be real. 
Non-diagonal entries in the lepton sector are expected to be generated when moving from the interaction basis to the 
mass basis. We will analyze this effect in Section~\ref{ssec:lepton-mixing}.

	\subsection{Mass basis}
	\label{ssec:mass-basis}

We define the rotation from flavor to mass eigenstates by unitary transformations as

\begin{align}
\begin{split}
u_L^\prime &= U_{L,u}\, u_L\,,\,\qquad d_L^\prime = U_{L,d}\, d_L\,,\,\qquad e_L^\prime = U_{L,\ell}\, e_L\,,\qquad \nu_L^\prime = U_{L,\ell}\, \nu_L\,,\\
u_R^\prime &= U_{R,u}\, u_R\,,\qquad d_R^\prime = U_{R,d}\, d_R\,,\qquad e_R^\prime = U_{R,\ell}\, e_R
\end{split}
\end{align}

\noindent where we recognize the CKM matrix $V =  U_{L,u}^\dagger U_{L,d}$, and neutrino masses have been neglected. To express the Lagrangian \eqref{eq:LNP0} in  the mass basis, it is convenient to define the matrices
\begin{align}
\label{eq:lambda-matrices}
\begin{split}
\lambda^{uu} &= U_{L,u}^\dagger \, P_3 \, U_{R,u}\,,\qquad\quad \lambda^{ud} = U_{L,u}^\dagger \, P_3 \, U_{R,d}\,,\\[0.3em]
\lambda^{du} &= U_{L,d}^\dagger \, P_3 \, U_{R,u}\,,\qquad\quad
 \lambda^{dd} = U_{L,d}^\dagger \, P_3 \, U_{R,d}\,,
\end{split}
\end{align}
and, similarly,
\begin{align}
\label{eq:lambda-matrices-lep}
\lambda^{\ell\,(k)} &= U_{L,\ell}^\dagger \, P_k \, U_{R,\ell}
\end{align}
\noindent where $(P_k)_{ij}=\delta_{ik}\delta_{jk}$, with $k\in\lbrace 1,2,3 \rbrace$, are projectors. These matrices are related to $V\equiv V_{CKM}$ via the relations $\lambda^{uu} = V \, \lambda^{du}$ and $\lambda^{dd} = V^\dagger \, \lambda^{ud}$. With these definitions we can express Eq.~\eqref{eq:LNP0} in the mass basis as
\begin{align}
\label{eq:LNP0-mass}
\mathcal{L}_{\mathrm{NP}}^0 = \sum_k &\Bigg{\lbrace} \dfrac{C_{S_R}^k}{\Lambda^2}\,\lambda^{\ell\, (k)}_{ij}\Big{[}\left(\lambda_{ts}^{ud}\right)^\ast \big{(}\bar{d}_{sR} u_{tL}\big{)}\big{(}\bar{\nu}_{iL} e_{jR}\big{)}+\left(\lambda_{ts}^{dd}\right)^\ast \big{(}\bar{d}_{sR} d_{tL}\big{)}\big{(}\bar{e}_{iL} e_{jR}\big{)}\Big{]} \nonumber\\[0.1em]
&+\dfrac{C_{S_L}^k}{\Lambda^2}\,\lambda^{\ell\, (k)}_{ij}\Big{[}\lambda_{st}^{du} \big{(}\bar{d}_{sL} u_{tR}\big{)}\big{(}\bar{\nu}_{iL} e_{jR}\big{)}-\lambda_{st}^{uu} \big{(}\bar{u}_{sL} u_{tR}\big{)}\big{(}\bar{e}_{iL} e_{jR}\big{)}\Big{]}\\[0.1em]
&+\dfrac{C_{T}^k}{\Lambda^2}\,\lambda^{\ell\, (k)}_{ij}\Big{[}\lambda_{st}^{du} \big{(}\bar{d}_{sL} \sigma^{\mu\nu}u_{tR}\big{)}\big{(}\bar{\nu}_{iL} \sigma_{\mu\nu} e_{jR}\big{)}-\lambda_{st}^{uu} \big{(}\bar{u}_{sL}  \sigma^{\mu\nu} u_{tR}\big{)}\big{(}\bar{e}_{iL}  \sigma_{\mu\nu}e_{jR}\big{)}\Big{]}\Bigg{\rbrace}+\mathrm{h.c.}\,,\nonumber
\end{align}

\noindent where the summation over the leptonic indices $i,j=e,\mu,\tau$ is implicit.

	\subsection{RGE flow above the electroweak scale}
	\label{ssec:rge}

We included the RGE electroweak effects for the Lagrangian~\eqref{eq:LNP0} at one-loop order in the leading logarithmic approximation, assuming that $y_t$ is the only relevant Yukawa coupling, c.f.~Ref.~\cite{Feruglio:2016gvd,Feruglio:2017rjo,Cornella:2018tfd} for details. By using the anomalous dimension matrices computed in Ref.~\cite{Jenkins:2013zja,Alonso:2013hga,Jenkins:2013wua}, which we have explicitly verified, we found that the Lagrangian at a scale $m_{\mathrm{EW}}< \mu <\Lambda$ now becomes $\mathcal{L}=\mathcal{L}_{\mathrm{SM}}+\mathcal{L}_{\mathrm{NP}}^0+\mathcal{L}_{\mathrm{eff}}$, where $\mathcal{L}_{\mathrm{eff}}$ denotes the RGE induced contributions. This term can be recast as
	
\begin{equation}
\label{eq:Leff-mZ}
\mathcal{L}_{\mathrm{eff}} = \delta \mathcal{L}_{\mathrm{SL}}+ \delta \mathcal{L}_{\mathrm{dip}}+ \delta \mathcal{L}_{\mathrm{H}}\,,
\end{equation}	

\noindent where $\delta\mathcal{L}_{\mathrm{SL}}$ denotes the loop corrections to the semileptonic operators of Eq.~\eqref{eq:LNP0-mass}, 
while $ \delta \mathcal{L}_{\mathrm{dip}}$ and $\delta \mathcal{L}_{\mathrm{H}}$ are new contributions to magnetic dipoles and to the Higgs interactions with charged leptons, respectively. In the mass basis, $\delta \mathcal{L}_{\mathrm{SL}}$ reads 

\begin{equation}
\label{eq:LSL-loop}
\delta \mathcal{L}_{\mathrm{SL}} = \dfrac{1}{16 \pi^2 \Lambda^2}\log \left(\dfrac{\Lambda}{\mu}\right) \sum_i \xi_i^{\mathrm{SL}} Q_i^{\mathrm{SL}}\,,
\end{equation}

\noindent where the operators $Q_i^{\mathrm{SL}}$ and the corresponding Wilson coefficients $\xi_i^{\mathrm{SL}}$ are collected in 
Table~\ref{tab:ope-SL-1}. An interesting feature that will be discussed below is the presence of a non-negligible 
electroweak mixing of the tensor coefficient $C_T^k$ into $C_{S_L}^k$, which is due to an accidentally large anomalous dimension, 
as pointed out in Ref.~\cite{Gonzalez-Alonso:2017iyc,Aebischer:2017gaw}.

The last two terms of Eq.~(\ref{eq:Leff-mZ}), $\delta \mathcal{L}_{\mathrm{dip}}$ and $\delta \mathcal{L}_{H}$, whose implications are the focus of our work, can be expressed in the mass basis as
\begin{align}
\label{eq:dip}
\delta \mathcal{L}_{\mathrm{dip}} &= \dfrac{\log \left(\Lambda/\mu\right)}{16 \pi^2 \Lambda^2}\sum_{k=e,\mu,\tau} C_T^k \, \lambda_{ij}^{\ell(k)}\,\lambda_{33}^{uu} \,y_t\,\left(6\, g_2 \big{[}\mathcal{O}_{eW}\big{]}_{ij}-10\, g_1 \big{[}\mathcal{O}_{eB}\big{]}_{ij} \right)+\mathrm{h.c.}\,,\\[0.4em]
\label{eq:Higgs}
\delta \mathcal{L}_{H} &= \dfrac{\log \left(\Lambda/\mu\right)}{16 \pi^2 \Lambda^2}\sum_{k=e,\mu,\tau} C_{S_L}^k \, \lambda_{ij}^{\ell(k)}\,\lambda_{33}^{uu}\, 12\,y_t\,(\lambda-y_t^2) \big{[}\mathcal{O}_{eH}\big{]}_{ij}+\mathrm{h.c.}\,,
\end{align}
	
\noindent where $\lambda$ is the Higgs quartic coupling $\lambda=m_H^2/(2v^2)$ and the effective dipole and Higgs operators are defined by 
\begin{align}
\big{[}\mathcal{O}_{eB}\big{]}_{ij} &= \big{(}\bar{L}_i\sigma^{\mu\nu}e_{jR}\big{)} H B_{\mu\nu}\,,\\[0.4em]
\big{[}\mathcal{O}_{eW}\big{]}_{ij} &= \big{(}\bar{L}_i\sigma^{\mu\nu}e_{jR}\big{)} \tau^I  H\, W_{\mu\nu}^I\,,\\[0.4em]
\big{[}\mathcal{O}_{eH}\big{]}_{ij} &= \big{(}H^\dagger H \big{)}\big{(}\bar{L}_i e_{jR}H\big{)}\,.
\end{align} 

\begin{table}[htbp!]
\renewcommand{\arraystretch}{2.8}
\centering
\begin{tabular}{|c|c|}\hline
$Q_i^{\mathrm{SL}}$  & $\xi_i^{\mathrm{SL}}$\\ \hline\hline 
$\big{(}\bar{d}_{sL} \sigma^{\mu\nu}u_{tR}\big{)}\big{(}\bar{\nu}_{Li} \sigma_{\mu\nu} e_{jR}\big{)}$	 & 	\specialcell{$\displaystyle\sum_k  \lambda_{ij}^{\ell(k)}\Bigg{\lbrace}\lambda_{st}^{du}\Bigg{[} C_T^k \left(-\dfrac{2 g_1^2}{9}+3 g_2^2-\dfrac{8g_3^2}{3}\right)-C_{S_L}^k \left(\dfrac{5g_1^2}{8}+\dfrac{3 g_2^2}{8}\right)\Bigg{]}$\\$- \, C_T^k \,y_t^2\Bigg{(}\dfrac{V_{s3}^\dagger\, \lambda^{uu}_{3t}}{2}+\lambda^{du}_{s3}\,\delta_{3t}\Bigg{)} \Bigg{\rbrace}$} \\ \hline
$\big{(}\bar{d}_{sL} u_{tR}\big{)}\big{(}\bar{\nu}_{Li} e_{jR}\big{)}$ & \specialcell{$\displaystyle\sum_k  \lambda_{ij}^{\ell(k)}\Bigg{\lbrace} \lambda_{st}^{du}\Bigg{[} C_{S_L}^k \left(\dfrac{11 g_1^2}{3}+8 g_3^2\right)-C_{T}^k \left(30 g_1^2+18g_2^2\right)\Bigg{]}$\\$+ C_{S_L}^k\,  y_t^2\Bigg{(}\dfrac{V^\dagger_{s3}\, \lambda^{uu}_{3t}}{2}+\lambda^{du}_{s3}\,\delta_{3t}\Bigg{)} \Bigg{\rbrace}$}\\ \hline 
$\big{(}\bar{d}_{sR} u_{tL}\big{)}\big{(}\bar{\nu}_{iL} e_{jR}\big{)}$ & $\displaystyle\sum_k \lambda_{ji}^{\ell(k)}\Bigg{[} \left(\lambda_{st}^{ud}\right)^\ast C_{S_R}^k \left(\dfrac{8 g_1^2}{3}+8 g_3^2\right)-  \left(\lambda^{ud}_{3s}\right)^\ast\, C_{S_R}^k \, \delta_{3t} \, \dfrac{y_t^2}{2}  \Bigg{]}$ \\ \hline
$\big{(}\bar{u}_{sL} \sigma^{\mu\nu}u_{tR}\big{)}\big{(}\bar{e}_{iL} \sigma_{\mu\nu} e_{jR}\big{)}$	& 	\specialcell{$\displaystyle\sum_k  \lambda_{ij}^{\ell(k)}\Bigg{\lbrace}\lambda_{st}^{dd}\Bigg{[}- C_T^k \left(-\dfrac{2 g_1^2}{9}+3 g_2^2-\dfrac{8g_3^2}{3}\right)-C_{S_L}^k \left(\dfrac{5g_1^2}{8}+\dfrac{3 g_2^2}{8}\right)\Bigg{]}$\\$+ \, C_T^k \,y_t^2\Bigg{(}\dfrac{\delta_{s3} \, \lambda^{uu}_{3t}}{2}+\lambda^{du}_{s3}\, \delta_{3t}\Bigg{)} \Bigg{\rbrace}$} \\ \hline
$\big{(}\bar{u}_{sL} u_{tR}\big{)}\big{(}\bar{e}_{iL} e_{jR}\big{)}$ & \specialcell{$\displaystyle\sum_k  \lambda_{ij}^{\ell(k)}\Bigg{\lbrace} \lambda_{st}^{dd}\Bigg{[} -C_{S_L}^k \left(\dfrac{11 g_1^2}{3}+8 g_3^2\right)+C_{T}^k \left(30 g_1^2+18g_2^2\right)\Bigg{]}$\\$+ C_{S_L}^k\,  y_t^2\Bigg{(}\dfrac{\delta_{s3} \ \lambda^{uu}_{3t}}{2}+\lambda^{du}_{s3} \, \delta_{3t}\Bigg{)} \Bigg{\rbrace}$} \\ \hline 
$\big{(}\bar{d}_{sR} d_{tL}\big{)}\big{(}\bar{e}_{iL} e_{jR}\big{)}$ & $\displaystyle\sum_k \lambda_{ij}^{\ell(k)}\Bigg{[} \left(\lambda_{st}^{uu}\right)^\ast C_{S_R}^k \left(\dfrac{8 g_1^2}{3}+8 g_3^2\right)-  \left(\lambda^{ud}_{3s}\right)^\ast\, C_{S_R}^k \, V_{3t} \, \dfrac{y_t^2}{2} \Bigg{]}$ \\ \hline
\end{tabular}
\caption{ \sl \small Semileptonic operators $Q_i^{S_L}$ and the corresponding effective coefficients $\xi_i^{\mathrm{SL}}$, c.f.~Eq.~\eqref{eq:LSL-loop}.}
\label{tab:ope-SL-1} 
\end{table}

\begin{figure}[ht!]
\centering
\includegraphics[width=0.92\linewidth]{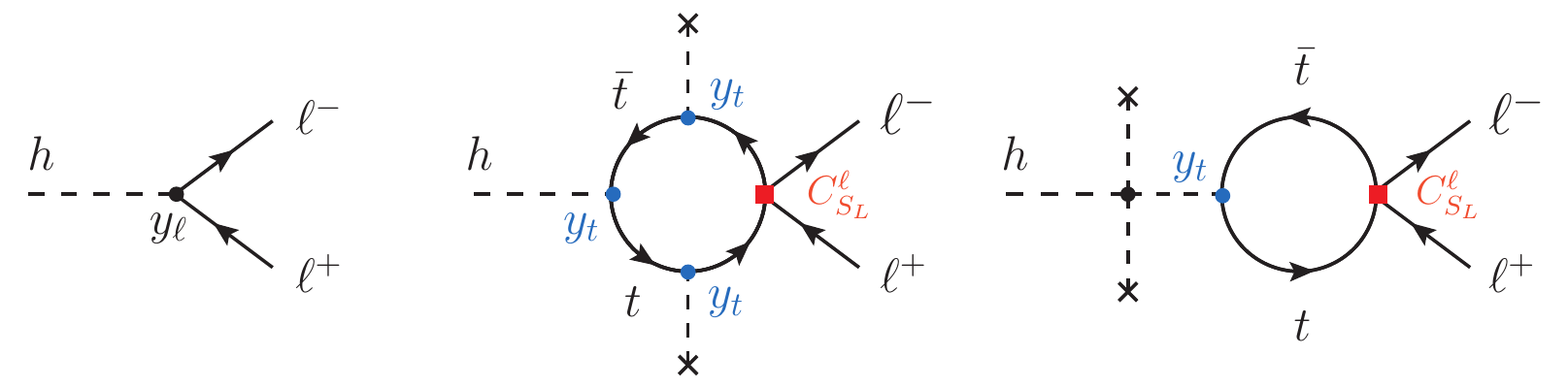}
\caption{\small \sl Higgs coupling to charged leptons in the SM and the new contributions generated by the Lagrangian~\eqref{eq:LNP0} through loop effects. While the SM term is proportional to the lepton Yukawas ($y_\ell$), the NP contributions are proportional to different powers of the top Yukawa ($y_t$), c.f.~Eq.~\eqref{eq:Higgs}. }
\label{fig:diagram-higgs}
\end{figure}

\begin{figure}[ht!]
\centering
\includegraphics[width=0.8\linewidth]{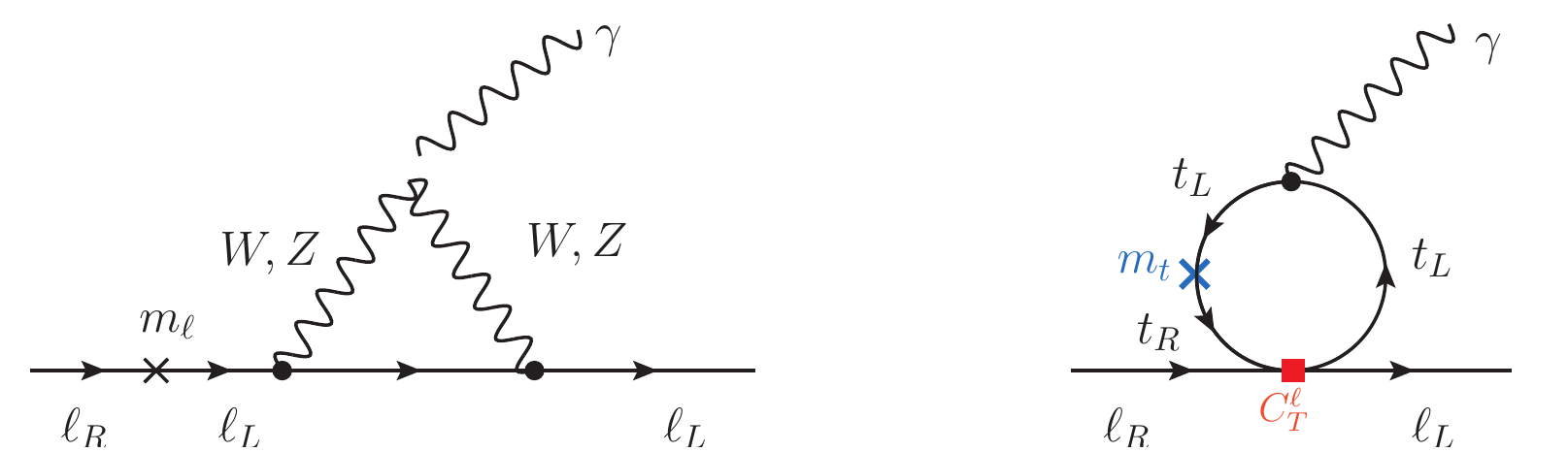}
\caption{\small \sl One sample of electroweak contributions to lepton dipoles in the SM (left), compared to the NP contribution induced 
by the Lagrangian~\eqref{eq:LNP0} at loop-level (right). The factor $m_\ell$ of the SM rates is replaced by $m_t$, thus compensating 
the loop suppression and inducing a large contribution to leptonic dipoles.}
\label{fig:diagram-dipole}
\end{figure}

\noindent Interestingly, the contributions in Eq.~\eqref{eq:dip} and~\eqref{eq:Higgs} are enhanced by $y_t$ due to the chiral structure of the operators $\mathcal{O}_{lequ}^{(1)}$ and $\mathcal{O}_{lequ}^{(3)}$. This feature is illustrated in Fig.~\ref{fig:diagram-higgs} and \ref{fig:diagram-dipole}, where we compare the new contributions with the SM ones for leptonic dipoles and charged lepton Yukawas, respectively. In both cases, the lepton helicity suppression of the SM contributions is lifted by the new operators, which receive a large contribution from the top quark. This is the main phenomenological feature of the scenario we consider, which will be further discussed in Sec.~\ref{ssec:magneticmoment} and~\ref{ssec:higgs}.

\section{Phenomenology}
\label{sec:pheno}

To proceed with our phenomenological analysis, we need additional assumptions on the mixing matrices feeding the NP effects
to the light generations. We will initially neglect the mixing effects in the leptonic sector by assuming $U_{L,\ell}=U_{R,\ell}=\mathbb{1}$, 
which implies that~$\lambda^{\ell(k)}=P_k$ in Eq.~\eqref{eq:lambda-matrices-lep}. As we will see in Section 3.5, non-diagonal 
entries in $U_{L,\ell}$ and $U_{R,\ell}$ are bound to be very small. Furthermore, we choose $U_{L,d}=\mathbb{1}$, implying that 
$V_{CKM}=U_{L,u}^\dagger$, and we assume the following structure for the transformation of right-handed fields:
\begin{align}
U_{R,u} = \begin{pmatrix}
0 & 0 & 0\\ 
0 & \cos \theta_U & -\sin \theta_U\\ 
0 & \sin \theta_U & \cos \theta_U 
\end{pmatrix}\,,
\qquad\qquad U_{R,d} =\mathbb{1} \,,
\end{align}

\noindent where $\theta_U$ is a generic angle. This set of assumptions on the quark mixing is minimal. With our choice of quark mixing the matrices defined in Eq.~\eqref{eq:lambda-matrices} read

\begin{align}
\label{eq:lambda-matrices-pheno}
\lambda^{du} &= \begin{pmatrix}
0 & 0 & 0\\ 
0 & 0 & 0\\ 
0 & \sin \theta_U & \cos \theta_U 
\end{pmatrix} \,,\qquad
\lambda^{uu} =\begin{pmatrix}
0 & \sin \theta_U\,V_{ub} & \cos \theta_U\,V_{ub}\\ 
0 & \sin \theta_U\,V_{cb} & \cos \theta_U\,V_{cb}\\ 
0 & \sin \theta_U\,V_{tb} & \cos \theta_U\,V_{tb} 
\end{pmatrix} \,,
\end{align}

\noindent while $\lambda^{dd}_{ij}= \delta_{3i}\delta_{3j}$ and $\lambda^{ud}_{ij}=\delta_{3j}\, V^\ast_{i3}$, providing the required couplings 
$\lambda_{32}^{du}$ and $\big{(}\lambda_{23}^{ud}\big{)}^\ast$ in a minimal form.

\

In the following we discuss the relevant observables for the scenario described above.

	\subsection{$R_D$ and $R_{D^\ast}$}
	\label{ssec:RD}
	
The first observables we consider are the lepton flavor universality ratios $R_D$ and $R_{D^\ast}$, defined by

\begin{equation}
\label{eq:RD-RDst}
R_{D^{(\ast)}} = \dfrac{\mathcal{B}(B\to D^{(\ast)}\tau \bar{\nu})}{\mathcal{B}(B\to D^{(\ast)}\ell \bar{\nu})}\,,\qquad\qquad \ell =e,\mu\,,
\end{equation}

\noindent which were found to be in disagreement with the SM predictions at the $B$--physics experiments~\cite{Ciezarek:2017yzh}. 
In particular,  the current experimental averages read~\cite{Amhis:2016xyh}
\begin{align}
R_D^{\mathrm{exp}}=0.41(5)\,,\qquad\qquad\quad R_{D^{\ast}}^{\mathrm{exp}}=0.304(15)\,,
\label{RD_RDstar_exp}
\end{align}

\noindent to be compared with our SM predictions
\begin{align}
R_D^{\mathrm{SM}}=0.293(7)\,,\qquad\qquad\quad R_{D^\ast}^{\mathrm{SM}}=0.257(3)\,,
\label{RD_RDstar_th}
\end{align}

\noindent which are respectively $\approx 2\,\sigma$ and $\approx 3\,\sigma$ larger than the experimental values given above.~\footnote{These predictions agree with other results in the literature, such as $R_D^{\mathrm{SM}}=0.299(3)$ \cite{Bigi:2016mdz}, $R_{D^\ast}^{\mathrm{SM}}=0.252(3)$~\cite{Fajfer:2012vx},
 $R_{D^\ast}^{\mathrm{SM}}=0.260(8)$~\cite{Bigi:2017jbd,Jaiswal:2017rve}.} The value for $R_D^{\mathrm{SM}}$ has been obtained by using the scalar and vector form factors computed by means of numerical simulations 
of QCD on the lattice (LQCD) in Ref.~\cite{Na:2015kha,Lattice:2015rga}, as well as the tensor one computed in Ref.~\cite{Atoui:2013mqa}. 
On the other hand, since the $B\to D^\ast$ form factors at nonzero recoil are not yet available from LQCD simulations, we consider the ones extracted from experimental results~\cite{Amhis:2016xyh}, combined with the ratios $A_0(q^2)/A_1(q^2)$ and $T_{1-3}(q^2)/A_1(q^2)$ computed in Ref.~\cite{Bernlochner:2017jka}. 

NP contributions for the transition $b\to c e_i \bar{\nu}_i$ can be described in full generality by the following dimension-6 effective Lagrangian 
at the scale $\mu=m_b$

\begin{align}
\label{leff:bc}
\mathcal{L}_{\mathrm{eff}} = -2 \sqrt{2}& G_F V_{cb} \bigg{[}(1+g_{V_L}^{i})\,\big{(}\bar{c}_L \gamma^\mu b_L\big{)}\big{(}\bar{e}_{iL} \gamma_\mu \nu_{iL}\big{)} +g_{V_R}^{i}\,\big{(}\bar{c}_R \gamma^\mu b_R\big{)}\big{(}\bar{e}_{iL} \gamma_\mu \nu_{iL}\big{)}\\
& + g_{S_R}^{i}\,\big{(}\bar{c}_L  b_R\big{)}\big{(}\bar{e}_{iR} \nu_{iL}\big{)}+ g_{S_L}^{i}\,\big{(}\bar{c}_R b_L\big{)}\big{(}\bar{e}_{iR} \nu_{iL}\big{)}+ g_T^{i}\, \big{(}\bar{c}_R \sigma^{\mu\nu} b_L\big{)}\big{(}\bar{e}_{iR} \sigma_{\mu\nu} \nu_{iL}\big{)}\bigg{]}+\mathrm{h.c.,}\nonumber
\end{align}

\noindent where the coefficients $g_{V_{L(R)}}^i$, $g_{S_{L(R)}}^i$ and $g_T^i$ are free parameters, and $i=e,\mu,\tau$, as before.~\footnote{Note that $g_{V_R}^\ell$ is generated by 
the dimension-6 operator $\mathcal{O}_{H u}= \big{(}H^\dagger i D_\mu H\big{)}\big{(}\bar{u}_{pR} \gamma^\mu u_{rR}\big{)}$ which does not break LFU~\cite{Fajfer:2012jt,Aebischer:2015fzz,Bernard:2006gy}. Moreover, it can be shown that the operator $(\bar{c}_L \sigma^{\mu\nu} b_R)(\bar{e}_R \sigma_{\mu\nu} \nu_L)$ vanishes because of the identity $\sigma^{\mu\nu}\gamma_5=\frac{i}{2}\epsilon^{\mu\nu\rho\sigma}\sigma_{\rho\sigma}$.} In the following we neglect the effective coefficients $g_{V_L}$ and $g_{V_R}$. The scenario considered in Eq.~\eqref{eq:LNP0} can be matched onto Eq.~\eqref{leff:bc} at the scale $\mu=m_b$ via the expression

\begin{align}
\label{eq:matching-gi-Ci}
\begin{split}
g_{S_L}^i (m_b) &= -\dfrac{ C_{S_L}^i(m_b)}{2\, V_{cb}} \dfrac{v^2}{\Lambda^2} \big{(}\lambda^{du}_{32}\big{)}^\ast + \dots\,,\qquad\quad
g_{T}^i (m_b) = -\dfrac{ C_{T}^i(m_b)}{2\, V_{cb}} \dfrac{v^2}{\Lambda^2} \big{(}\lambda^{du}_{32}\big{)}^\ast+\dots\,,
\end{split}
\end{align}
\noindent where $\lambda_{32}^{du}= \sin{\theta_U}$, while for the other Wilson coefficient
\begin{align}
\label{eq:matching-gi-Ci-bis}
\begin{split}
g_{S_R}^i (m_b) &= -\dfrac{ C_{S_R}^i(m_b)}{2\, V_{cb}} \dfrac{v^2}{\Lambda^2} \lambda^{ud}_{23}+\dots\,,
\end{split}
\end{align}

\noindent where $\lambda^{ud}_{23}= V_{cb}^\ast$. Dots in the above equations stand for RGE-induced contributions, c.f.~Eq.\eqref{eq:LSL-loop}.  In the following, we will consider two representative NP scenarios: 

\begin{itemize}
	\item[$(i)$] Scalar and tensor operators containing left-handed down-quarks: $C_{S_L}^\tau,~C_{T}^\tau\neq 0$;
	\item[$(ii)$] Scalar operators: $C_{S_L}^\tau,~C_{S_R}^\tau\neq 0$,
\end{itemize}

\noindent with the other Wilson coefficients set to zero.  Differently from the (current)$\times$(current) operators, which are invariant under QCD running, the scalar and tensor operators are renormalized by strong interactions. These effects from $\mu=1$~TeV down to $\mu=m_b$ have been computed at one-loop in QED/EW and three-loops in QCD in Ref.~\cite{Gonzalez-Alonso:2017iyc}. We recall that there is a large mixing of $C_T^\tau$ into $C_{S_L}^\tau$, but not the other way around. Moreover, $C_{S_R}^\tau$ has a negligible 
mixing into the other Wilson coefficients. For this reason, the above choice of effective scenarios $(i)$ and $(ii)$ is stable under the RGE flow. 

To identify the range of $C_{S_{L(R)}}^\tau$ and $C_T^\tau$ favored by the $b\to c$ anomalies, we perform a fit to the current experimental averages of $R_{D}$ and $R_{D^{\ast}}$~\cite{Amhis:2016xyh}. Compact 
expressions for these observables can be obtained by using the hadronic inputs described above and the decay rate expressions from Ref.~\cite{Becirevic:2016oho,Sakaki:2013bfa}. By assuming that NP only contributes to the transition 
$b\to c \tau \bar{\nu}$, we obtain that

\begin{align}
\label{eq:RD-RDst-compact}
\begin{split}
\dfrac{R_{D^{(\ast)}}}{R_{D^{(\ast)}}^{\mathrm{SM}}} =  1 &+ a^{D^{(\ast)}}_{S} \, |g_S^\tau|^2 + a^{D^{(\ast)}}_{P} \, |g_P^\tau|^2+ a_T^{D^{(\ast)}} \, |g_T^\tau|^2 \\[0.3em]
&+ a^{D^{(\ast)}}_{SV_L}\,\mathrm{Re}\left[g_S^\tau\right]+ a^{D^{(\ast)}}_{PV_L}\,\mathrm{Re}\left[g_P^\tau\right]+ a_{TV_L}^{D^{(\ast)}}\,\mathrm{Re}\left[g_T^\tau\right]\,,
\end{split}
\end{align}

\noindent where $g_{S(P)}^\tau = g_{S_R}^\tau \pm g_{S_L}^\tau$, and the numeric coefficients $a_i^D$ and $a_i^{D^\ast}$ are collected in 
Table~\ref{tab:RD-RDst-compact}. Note that these coefficients have nontrivial correlations which are taken into account in our 
numerical analysis, even though they are not given in Table~\ref{tab:RD-RDst-compact}. Interestingly, $R_D$ and $R_{D^\ast}$ are sensitive to a complementary set of NP operators. While $R_D$ is sensitive to both scalar and tensor contributions, $R_{D^\ast}$ only receives sizable contributions from the tensor ones.

\

\begin{table}[htbp!]
\renewcommand{\arraystretch}{1.5}
\centering
\begin{tabular}{|c|cccccc|}\hline
Decay mode  & $a_S^M$ & $a_{SV_L}^M$ & $a_P^M$ & $a_{PV_L}^M$ & $a_T^M$ & $a_{TV_L}^M$\\ \hline\hline 
$B \to D$	&	 $1.08(1)$   &     $1.54(2)$      &  $0$	&	$0$ & $0.83(5)$	& $1.09(3)$\\
$B\to {D^\ast}$	&	  $0$  &    $0$   & $0.0473(5)$	&	$0.14(2)$ &	$17.3(16)$ & $-5.1(4)$ \\ \hline
\end{tabular}
\caption{ \sl \small Numeric coefficients in Eq.~\eqref{eq:RD-RDst-compact} for $M=D, D^\ast$ obtained by using the hadronic parameters described in Sec.~\ref{ssec:RD}, assuming that NP only modifies the coefficients $g_{S_L}$, $g_{S_R}$ and $g_T$ for the transition $b\to c \tau \bar{\nu}$, c.f.~Eq.~\eqref{eq:matching-gi-Ci}. }
\label{tab:RD-RDst-compact} 
\end{table}

\begin{figure}[ht!]
\centering
\includegraphics[width=0.503\linewidth]{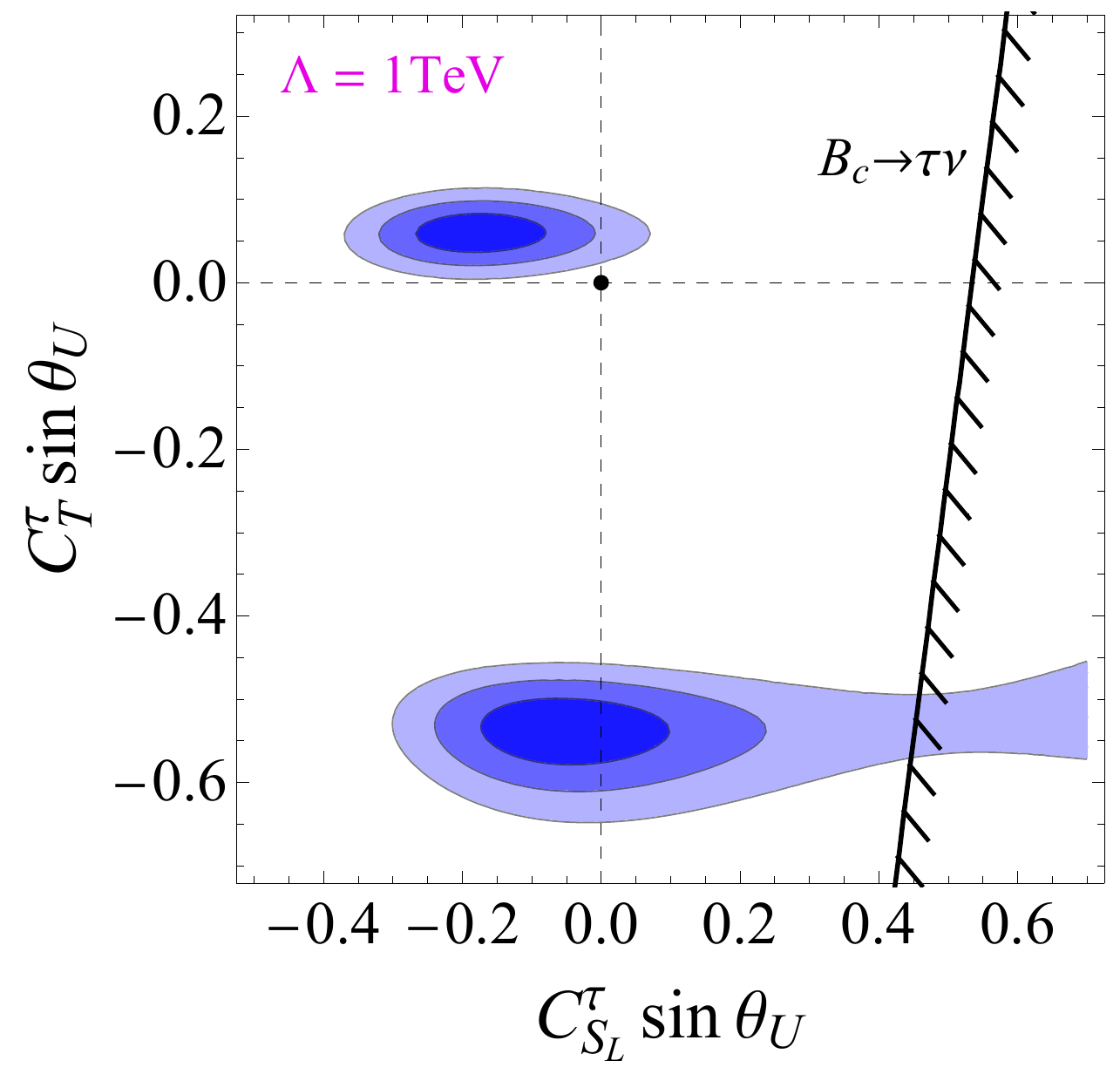}~\includegraphics[width=0.498\linewidth]{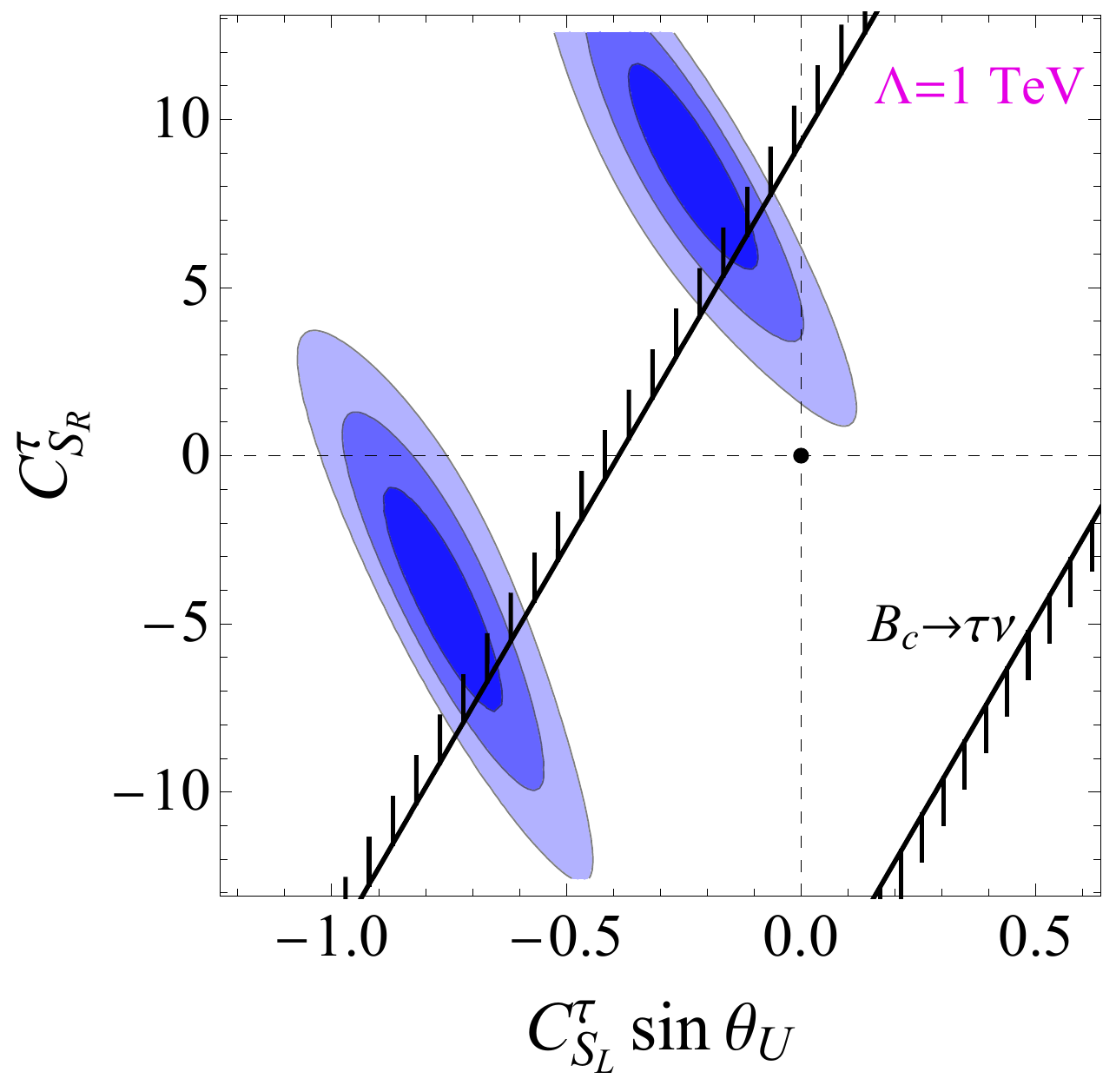}
\caption{\small \sl Allowed regions by $R_D$ and $R_{D^\ast}$ in the plane $C_{S_L}^\tau \, \sin \theta_U$ vs.~$C_{T}^\tau\, \sin \theta_U$ (left panel), 
and $C_{S_L}^\tau \sin \theta_U$ vs.~$C_{S_R}^\tau$ (right panel) are shown to $1$, $2$ and $3\,\sigma$ accuracy in blue (darker to lighter). 
We accounted for the electroweak and QCD running from the NP scale $\Lambda \approx 1~\mathrm{TeV}$ down to $\mu =m_b$, where 
the hadronic parameters are computed, c.f.~Ref.~\cite{Gonzalez-Alonso:2017iyc}. The black line shows the constraint from the $B_c$--meson lifetime, which allows us to exclude the solutions with large values of $|g_P|$, as explained in the text.}
\label{fig:RD-RDst-1TeV}
\end{figure}

 Our results for both scenarios are shown in Fig.~\ref{fig:RD-RDst-1TeV} at the scale $\mu=1$~TeV.
On the same plot, we show the constraints from $\mathcal{B}(B_c\to \tau \bar{\nu})$, derived from the $B_c$--meson lifetime, which are 
particularly useful to constraint $g_P^\tau$~\cite{Alonso:2016oyd,Li:2016vvp,Celis:2016azn}. This can be understood from the lift of the helicity suppression of the SM rate by pseudoscalar operators, as it can be seen in
\begin{equation}
\mathcal{B}(B_c\to \tau \bar{\nu}) = \tau_{B_c} \dfrac{m_{B_c} f_{B_c}^2 G_F^2 |V_{cb}|^2}{8 \pi} m_\tau^2 \left( 1- \dfrac{m_\tau^2}{m_{B_c}^2} \right)^2 \Bigg{|} 1 + g_{P}^\tau \dfrac{m_{B_c}^2}{m_\tau (m_b+m_c)}\Bigg{|}^2\,,
\end{equation}
\noindent where $f_{B_c}=427(6)$~MeV is the $B_c$--meson decay constant~\cite{McNeile:2012qf}. The current experimental value on $\tau_{B_c}= 0.507(9)\times 10^{-12}$~s allows us to set a conservative limit of $30\%$ on $\mathcal{B}(B_c\to \tau \bar{\nu})$~\cite{Alonso:2016oyd}, which can then be translated onto the following $1\,\sigma$ bound~\footnote{Alternatively, one could consider the less conservative limit $\mathcal{B}(B_c \to \tau \bar{\nu}) \lesssim 10\%$ computed in Ref.~\cite{Akeroyd:2017mhr}. By using this constraint, the limit in Eq.~\eqref{eq:gP-Bclifetime} would become $g_P\in (-0.76,0.30)$ to $1\,\sigma$ accuracy.}

\begin{equation}
\label{eq:gP-Bclifetime}
g_P^\tau (\mu=m_b) \equiv g_{S_R}^\tau(m_b)-g_{S_L}^\tau(m_b)\in (-1.14,0.68)\,.
\end{equation}

\noindent where we have used $|V_{cb}|=0.0417(20)$~\cite{Bigi:2017njr,Grinstein:2017nlq}. From Fig.~\ref{fig:RD-RDst-1TeV} we see that this constraint creates a tension on the (pseudo)scalar scenario depicted in the right panel, but it is still not sufficient to affect the scenario shown in 
the left panel. Interestingly, there are two viable solutions, namely,
\begin{align}
\label{eq:benchmark}
\dfrac{\sin \theta_U}{\Lambda^2}\,(C_{S_L}^\tau,C_T^\tau) \approx (-0.08,-0.53)~\mathrm{TeV}^{-2}\quad \text{or} \quad (-0.16,0.07)~\mathrm{TeV}^{-2} \,.
\end{align}

\noindent The existence of two solutions, independently found in Refs.~\cite{Freytsis:2015qca,Bhattacharya:2018kig,Azatov:2018knx}, 
rely on the interplay between linear and quadratic NP contributions, as one can easily see from Eq.~(\ref{eq:RD-RDst-compact}). 

Finally, for completeness, we also comment on the effective couplings to electrons and muons. An important constraint on scenarios aiming to simultaneously explain the LFU anomalies arises from the LFU tests between the transitions $b\to c\mu\bar{\nu}$ and $b\to c e\bar{\nu}$~\cite{Becirevic:2016oho,Jung:2018lfu}. Belle collaboration reported the experimental result~\cite{Glattauer:2015teq}

\begin{equation}
R_D^{\mu/e} = \dfrac{\mathcal{B}(B\to D \mu \bar{\nu})}{\mathcal{B}(B\to D e \bar{\nu})} = 0.995 \pm 0.022 \pm 0.039\,,
\end{equation}

\noindent which allows us to set constraints on $C_{S_L}^\ell$ and $C_T^\ell$ with $\ell=e,\mu$. By neglecting the couplings to electrons, we obtain the following $2\,\sigma$ constraints at $\mu =1$~TeV, 

\begin{equation}
\sin \theta_U  \dfrac{C_{S_L}^\mu}{\Lambda^2} \in  (-0.18,0.28)~\mathrm{TeV}^{-2} \,, \qquad \sin \theta_U  \dfrac{C_{T}^\mu}{\Lambda^2} \in (-0.81,1.06)~\mathrm{TeV}^{-2} \,,
\end{equation}

\noindent where we have assumed $|C_{S_L}^\mu| \gg |C_{S_L}^e|$ and $|C_{T}^\mu| \gg |C_{T}^e|$. We do not perform a similar analysis for $R_{D^\ast}^{\mu/e} = 0.961(47)$~\cite{Abdesselam:2017kjf} because the relevant form factors at nonzero recoil values are not yet available 
from LQCD calculations.

	\subsection{Leptonic anomalous magnetic moments}
	\label{ssec:magneticmoment}

Another important hint of NP comes from the muon's anomalous magnetic moment. Current experimental measurement indicates a deviation from the SM of about $\approx 3.6 \,\sigma$ in the observable~\cite{Bennett:2006fi,Jegerlehner:2009ry,NYFFELER:2014pta}
\begin{equation}
\label{eq:gminus2-muon}
\Delta a_\mu = a_\mu^{\mathrm{exp}} - a_{\mu}^{\mathrm{SM}} = (2.8 \pm 0.9) \times 10^{-9}\,,
\end{equation}
where $a_\mu = (g_\mu-2)/2$. 
Interestingly, the observed discrepancy is of the same order as the SM electroweak effect~\cite{Altarelli:1972nc}:
\begin{equation}
\label{eq:gminus2-muon_SM}
(a^{\rm SM}_\ell)_{\rm EW} = \frac{m^2_\ell}{(4\pi v)^2}\left( 1 - \frac{4}{3}\sin^2\theta_W + \frac{8}{3}\sin^4\theta_W \right)
\approx 2\times 10^{-9} \, \frac{m^2_\ell}{m^2_\mu}\,.
\end{equation}
If NP appears at a scale $\Lambda \gtrsim 1$ TeV, its effect on $a^{\rm NP}_\ell$ is expected to be of order 
$a^{\rm NP}_\ell/(a^{\rm SM}_\ell)_{\rm EW} \sim v^2/\Lambda^2$ and therefore safely negligible unless some enhancement
mechanism takes place.

In the NP scenario we consider, the dipole operators in Eq.~\eqref{eq:dip} contribute to electromagnetic dipoles as
%
\begin{align}
\label{eq:dip-bis}
\delta \mathcal{L}_{\mathrm{dip}} = -\dfrac{\log \left(\Lambda/m_t\right)}{16 \pi^2 \Lambda^2}\sum_{k=e,\mu,\tau} C_T^k \, \lambda_{ij}^{\ell(k)}\,\lambda_{33}^{uu}\, e \,y_t\, 16  \big{(} \bar{e}_{iR} \sigma^{\mu\nu} e_{jR}\big{)}\dfrac{v+h}{\sqrt{2}} F_{\mu\nu}\,,
\end{align}
%
%
leading to the following NP contribution to $a^{\rm NP}_{\ell}$
%
\begin{equation}
\label{eq:amu-NP}
a^{\rm NP}_{\ell} = -\frac{m^2_{\ell}}{(4\pi v)^2}\, \frac{m_t}{m_\ell}  \left(\frac{v}{\Lambda}\right)^2
64 \,C_T^\ell\, \lambda_{33}^{uu}\, \log \left( \Lambda/m_t \right)\,,
\end{equation}
%
\noindent where we have used the fact that $\lambda^{\ell(k)}_{ij}=\delta_{ki}\delta_{kj}$ in our setup. 
As already anticipated, Eq.~(\ref{eq:amu-NP}) exhibits a chiral enhancement $m_t/m_\ell$ which can easily 
compensate the suppression factor $(v/\Lambda)^2$.
As a result, we obtain that the discrepancy in $(g-2)_\mu$ can be accommodated to $1\,\sigma$ accuracy if

\begin{equation}
\label{eq:benchmark-gminus2}
\cos \theta_U \dfrac{C_T^\mu}{\Lambda^2} \in(-2.9,-1.5)\times 10^{-4}~\mathrm{TeV}^{-2}\,, 
\end{equation}
\noindent where we have set $\lambda_{33}^{uu}=V_{tb} \cos\theta_U$ and $\Lambda\approx 1$~TeV in the logarithmic terms. For the electron $g-2$, a conservative bound of $|\Delta a_e| \lesssim 8 \times 10^{-13}$ is obtained by using atomic physics determinations of the fine structure constant~\cite{Giudice:2012ms}. This limit can then be translated into~\footnote{Another possibility is to require that the NP contribution should be smaller than the current experimental error on $a_e$, namely, $|\Delta a_e|<2.6 \times 10^{-16}$~\cite{Agashe:2014kda}. By using this constraint we obtain the limit $|\cos \theta_U \, C_T^e \vert/\Lambda^2 < 3 \times 10^{-6}~\mathrm{TeV}^{-2}$, which is an order of magnitude more restrictive than the one in Eq.~\eqref{eq:gminu2e-constraint}.} 

\begin{equation}
\label{eq:gminu2e-constraint}
\vert\cos \theta_U \vert\dfrac{\vert C_T^e \vert}{\Lambda^2} < 1.3 \times 10^{-5}~\mathrm{TeV}^{-2}\,, 
\end{equation}
\noindent which is one order of magnitude more stringent than the one derived above on $C_T^\mu$. This result can be 
traced back to the scaling of Eq.~(\ref{eq:amu-NP}), $a^{\rm NP}_\ell \propto m_\ell \, m_t$, 
which differs from the SM one, $a^{\rm SM}_\ell \propto m_\ell^2$. As a result, the muon $g-2$ can be explained only if the tensor 
couplings are hierarchical, namely, $|C_T^\mu| \gtrsim 10\, |C_T^e|$. Since the explanation of $R_{D^{(\ast)}}$ requires 
much larger couplings than the ones obtained above from $(g-2)_{e,\mu}$,  in the present framework a simultaneous 
explanation of these anomalies is possible only if

\begin{equation}
|C_T^\tau| \gg |C_T^\mu| \gtrsim 10\, |C_T^e|\,.
\end{equation}

\noindent This suggests the possibility of testing this scenario with the $\tau$-lepton $g-2$. In particular, 
by taking the two benchmark values in Eq.~\eqref{eq:benchmark}, we obtain that 
\begin{equation}
\Delta a_\tau \approx -4 \times 10^{-4} \left(\frac{V_{cb}}{\tan\theta_U}\right)\,,\qquad
\Delta a_\tau \approx 3 \times 10^{-3} \left(\frac{V_{cb}}{\tan\theta_U}\right)\,,
\end{equation}
which lies possibly within reach of future experiments. 
This is one of our main predictions, which will be quantified in Sec.~\ref{sec:numerical} along with our numerical results.
While the current experimental sensitivity provides only a modest bound $-0.007< a_\tau<0.005$~\cite{GonzalezSprinberg:2000mk}, 
there are proposals to improve this limit at Belle-II~\cite{Eidelman:2016aih}.

	\subsection{Higgs decays}		
	\label{ssec:higgs}

Another observable exhibiting the chiral enhancement $m_t/m_\ell$ is the Higgs decay width into charged leptons, as illustrated in Fig.~\ref{fig:diagram-higgs}. The NP contributions in Eq.~\eqref{eq:Higgs} induce the following modification of the Yukawa 
couplings to leptons

\begin{equation}
\label{eq:yuk-eff}
y_{ij}^{\mathrm{eff}} = y_i \,\delta_{ij}   - 12\, C_{S_L}^i \lambda_{33}^{uu} \, y_t \, (\lambda-y_t^2) \dfrac{v^2\log \left( \Lambda/m_t \right)}{16\pi^2 \Lambda^2}\delta_{ij}\,,
\end{equation}

\noindent where $y_i={\sqrt{2}m_{i}}{v}$, with $i,j=e,\mu,\tau$, and we have assumed no mixing in the leptonic sector, as before. These Yukawa couplings contribute to the $H\to \ell^+_i \ell^-_i$ decay widths as

\begin{equation}
\Gamma (h\to \ell^+_i\ell^-_i) = \dfrac{m_h}{16\pi}\, |y^{\mathrm{eff}}_{ii}|^2\,,
\end{equation}

\noindent where we neglected $m_{\ell_i}$ in the phase space factors. The most interesting channel for us is $h\to\tau\tau$, which is constrained 
by the signal strength measured at the LHC~\cite{Agashe:2014kda}, 

\begin{equation}
\label{eq:htautau-exp}
\mu_{\tau\tau}^{\mathrm{exp}} = \dfrac{\sigma \cdot \mathcal{B}(h\to\tau\tau)}{\sigma_{\mathrm{SM}} \cdot \mathcal{B}(h\to\tau\tau)_{\mathrm{SM}}} = 1.12(23)\,,
\end{equation}
%
\noindent where $\sigma$ denotes the Higgs production cross-section. By assuming that $\sigma_{\mathrm{SM}}$ is not 
modified by NP, we obtain
%
\begin{equation}
\mu_{\tau\tau} \simeq 
\left| 
1  - 12 \cos\theta_{U}\, C_{S_L}^\tau \, \frac{m_t}{m_\tau} \, (\lambda-y_t^2) \dfrac{v^2\log \left( \Lambda/m_t \right)}{16\pi^2 \Lambda^2}
\right|^2\,,
\end{equation}
%
and therefore we get the following $2\,\sigma$ constraint
%
\begin{equation}
\cos\theta_U \dfrac{C_{S_L}^\tau}{\Lambda^2} \in (-0.31,0.44)~\mathrm{TeV}^{-2}\,.
\end{equation}
%

\noindent For instance, the benchmark point of Eq.~(\ref{eq:benchmark}), $ \sin \theta_U\,  C_{S_L}^\tau/\Lambda^2 \approx -0.16~\mathrm{TeV}^{-2}$, implies that
%
\begin{equation}
\mu_{\tau\tau} \simeq
\left(1  - 2.5 \,\frac{V_{cb}}{\tan\theta_U} \right)^2\,.
\end{equation}
Therefore, by combining the experimental constraints on $R_{D^{(\ast)}}$, measured at low-energies, and $h\to \tau\tau$, obtained at the LHC, we are able to extract information on the flavor mixing angle $\theta_U$. Interestingly, even a not so precise measurement of $\mu_{\tau\tau}$ is already enough to provide us independent information due to the large chiral enhancement by the top mass in Eq.~\eqref{eq:yuk-eff}.

	\subsection{Lepton flavor violating decays}
	\label{ssec:lepton-mixing}

The constraints derived above assume an alignment between flavor and mass basis for charged leptons, i.e.~$U_{L\ell}=U_{R\ell}=\mathbb{1}$.
In order to quantify the maximum misalignment allowed by current data, we parameterize the matrices $U_{L\ell}$ and $U_{R\ell}$ in a small mixing approximation, namely,

\begin{align}
U_{L\ell} = \begin{pmatrix}
1 & v_{12} & v_{13} \\ 
-v_{12}^\ast & 1 & v_{23}\\ 
-v_{13}^\ast & -v_{23}^\ast & 1 
\end{pmatrix}\,,\qquad\quad 
U_{R\ell} = \begin{pmatrix}
1 & u_{12} & u_{13} \\ 
-u_{12}^\ast & 1 & u_{23}\\ 
-u_{13}^\ast & -u_{23}^\ast & 1 
\end{pmatrix}\,,
\end{align}

\noindent where $|v_{ij}|,|u_{ij}| \ll 1$, with $i,j=1,2,3$. In our setup, the most stringent constraints on the lepton mixing parameters stem from the lepton flavor violating (LFV) decays $\ell \to \ell^\prime \gamma$, with $\ell,\ell^\prime=e,\mu,\tau$. These processes are experimentally constrained by~\cite{Agashe:2014kda}
\begin{align}
\begin{split}
\mathcal{B}(\mu \to e \gamma)^\mathrm{exp} &< 4.2 \times 10^{-13}\,,\\[0.4em]
\mathcal{B}(\tau \to e \gamma)^\mathrm{exp} &< 3.3 \times 10^{-8}\,,\\[0.4em]
\mathcal{B}(\tau \to \mu \gamma)^\mathrm{exp} &< 4.4 \times 10^{-8}\,.
\end{split}
\end{align}

\noindent By assuming $C_{T}^e$=0 and $|C_T^\mu| \ll |C_T^\tau| $, as suggested by the discussion above, we extract the following constraints
\begin{align}
\begin{split}
\sqrt{|u_{12}|^2+|v_{12}|^2}\, \dfrac{|\cos \theta_U\,C_T^\mu|}{\Lambda^2} & \lesssim 2.0 \times 10^{-9}~\mathrm{TeV}^{-2}\,,\\[0.3em]
\sqrt{|u_{13}|^2+|v_{13}|^2}\, \dfrac{|\cos \theta_U\,C_T^\tau|}{\Lambda^2} & \lesssim 2.5 \times 10^{-5}~\mathrm{TeV}^{-2}\,,\\[0.3em]
\sqrt{|u_{23}|^2+|v_{23}|^2}\, \dfrac{|\cos \theta_U\,C_T^\tau|}{\Lambda^2}  & \lesssim 2.6 \times 10^{-5}~\mathrm{TeV}^{-2}\,,
\end{split}
\end{align}

\noindent where we kept only the linear terms on $u_{ij}$ and $v_{ij}$ at the amplitude level. The main message from these equations is 
that the mixing angles are constrained to be very small. Furthermore, the presence of nonzero mixing allows us to correlate observables 
such as $\mathcal{B}(\mu\to e\gamma)$ with the muon $g-2$. 
We find that
%
\begin{align}
\begin{split}
\mathcal{B}(\mu\to e\gamma) \approx 5 \times 10^{-13} \left( \frac{\Delta a_\mu}{3\times 10^{-9}}\right)^2
\left(\frac{\sqrt{|u_{12}|^2+|v_{12}|^2}}{10^{-5}}\right)^2\,,
\end{split}
\end{align}

\noindent which tells us that an explanation of the $(g-2)_\mu$ anomaly is possible only if $\sqrt{|u_{12}|^2+|v_{12}|^2} \lesssim 10^{-5}$. 
Similarly, $\mathcal{B}(\tau\to \mu\gamma)$ is correlated with $\Delta a_\tau$ as follow 
%
\begin{align}
\begin{split}
\mathcal{B}(\tau\to\mu\gamma) \approx 10^{-7} \left( \frac{\Delta a_\tau}{10^{-4}}\right)^2
\left(\frac{\sqrt{|u_{23}|^2+|v_{23}|^2}}{10^{-4}}\right)^2\,,
\end{split}
\end{align}

\noindent implying that values of $\Delta a_\tau$ larger than $\approx 10^{-4}$ are allowed only if 
$\sqrt{|u_{23}|^2+|v_{23}|^2} \lesssim 10^{-4}$.

\section{Numerical analysis}
\label{sec:numerical}
In this Section we quantify our predictions for the most relevant observables identified in Sec.~\ref{sec:pheno}, namely, the decay $h\to\tau\tau$ and the $\tau$-lepton $g-2$. To this purpose, we perform a scan of the parameters randomly chosen over the ranges

\begin{equation}
C_{S_L}^\tau\,, C_{T}^\tau \in (-\sqrt{4\pi},\sqrt{4\pi})\,,\qquad  C_{T}^\mu  \in (-0.1,0.1) \,, \qquad \theta_U \in (0,2\pi)\,,
\label{eq:scan}
\end{equation}

\noindent from which we select the points consistent with both $R_{D^{(\ast)}}$ and $(g-2)_\mu$ to $1\,\sigma$ accuracy. The couplings $C_{S_L}^\mu$, $C_{S_L}^e$ and $C_{T}^e$ are set to zero, since they play no role in the phenomenology we want to discuss. Furthermore, we neglect the misalignment between flavor and mass basis of leptons, since it is tightly constrained by LFV decays, as discussed above.~\footnote{Note that the upper bound on $|C_T^\mu|$ in Eq.~\eqref{eq:scan} is chosen in such a way that this coupling will not significantly modify the denominator of $R_{D^{(\ast)}}$, as tacitly assumed in our discussion in Sec.~\ref{ssec:RD}. } 

\begin{figure}[p!]
\centering
\includegraphics[width=0.48\linewidth]{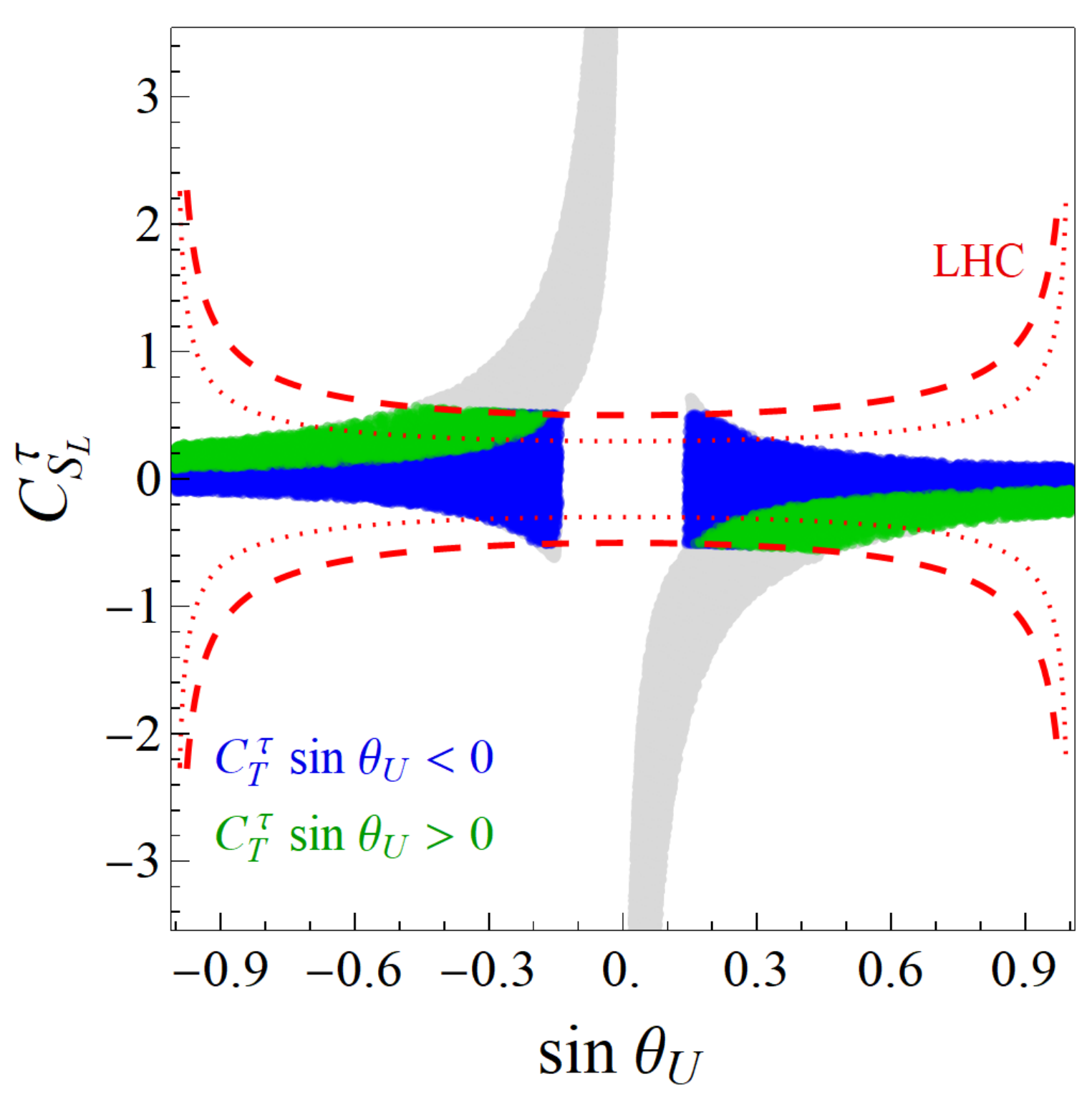}\quad\includegraphics[width=0.48\linewidth]{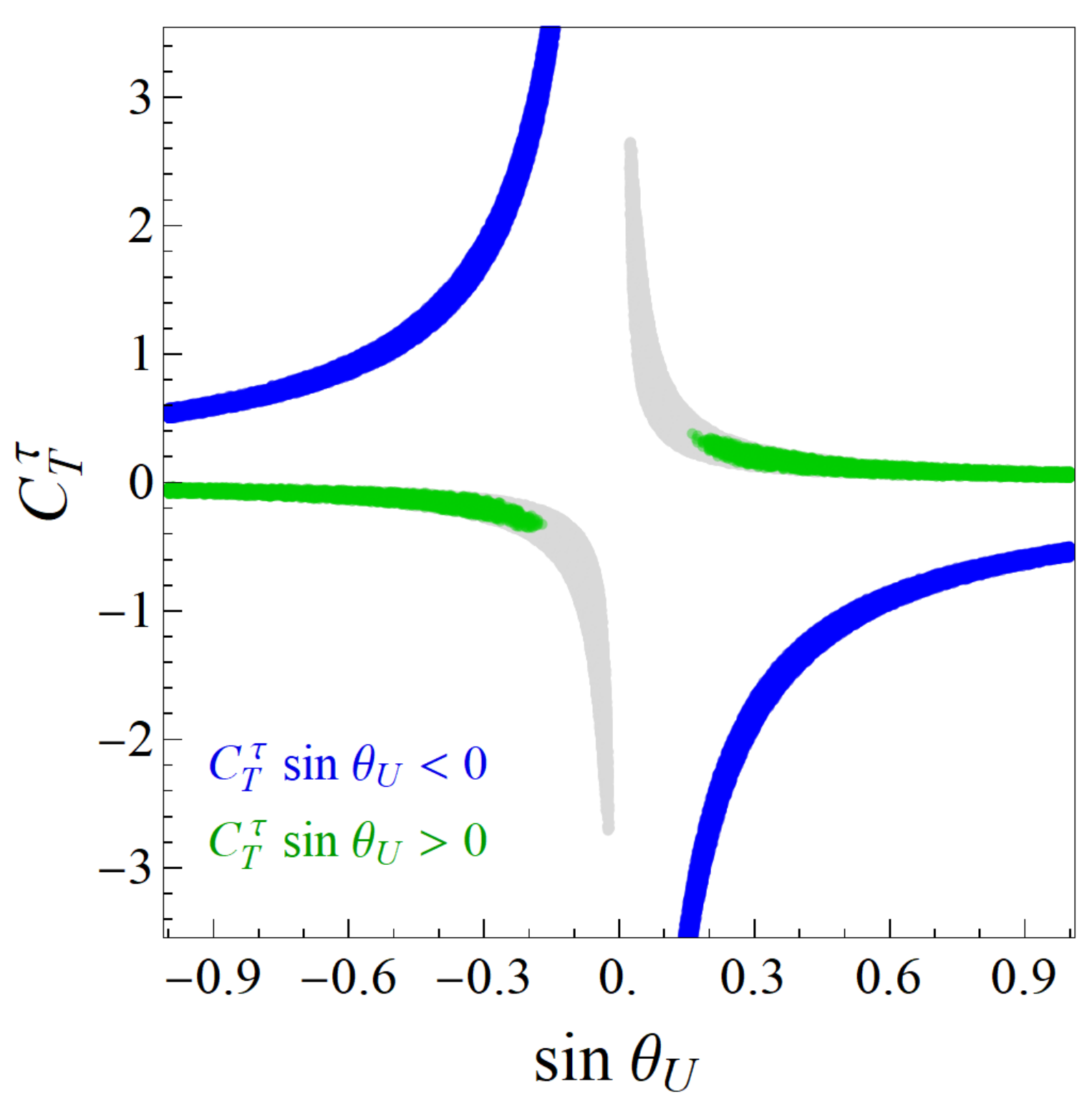}
\caption{\small \sl $\sin \theta_U$ is plotted against $C_{S_L}^\tau$ (left panel) and $C_T^{\tau}$ (right panel) for the points allowed by both $(g-2)_\mu$ and $R_{D^{(\ast)}}$ to $1\,\sigma$ accuracy. Gray points are excluded by the LHC results on the strength signal $\mu_{\tau\tau}$ [c.f.~Eq.~\eqref{eq:htautau-exp}], as illustrated by the red dotted ($1\,\sigma$) and dashed ($2\,\sigma$) lines on the left panel. The blue and green points are allowed by all observables for the two solutions found in Fig.~\ref{fig:RD-RDst-1TeV}, i.e.~the one with $C_{T}^\tau \,\sin \theta_U <0$ (blue) and the one with $C_T^\tau \,\sin \theta_U >0$ (green).}
\label{fig:results-couplings}
\end{figure}

\begin{figure}[p!]
\centering
\includegraphics[width=0.52\linewidth]{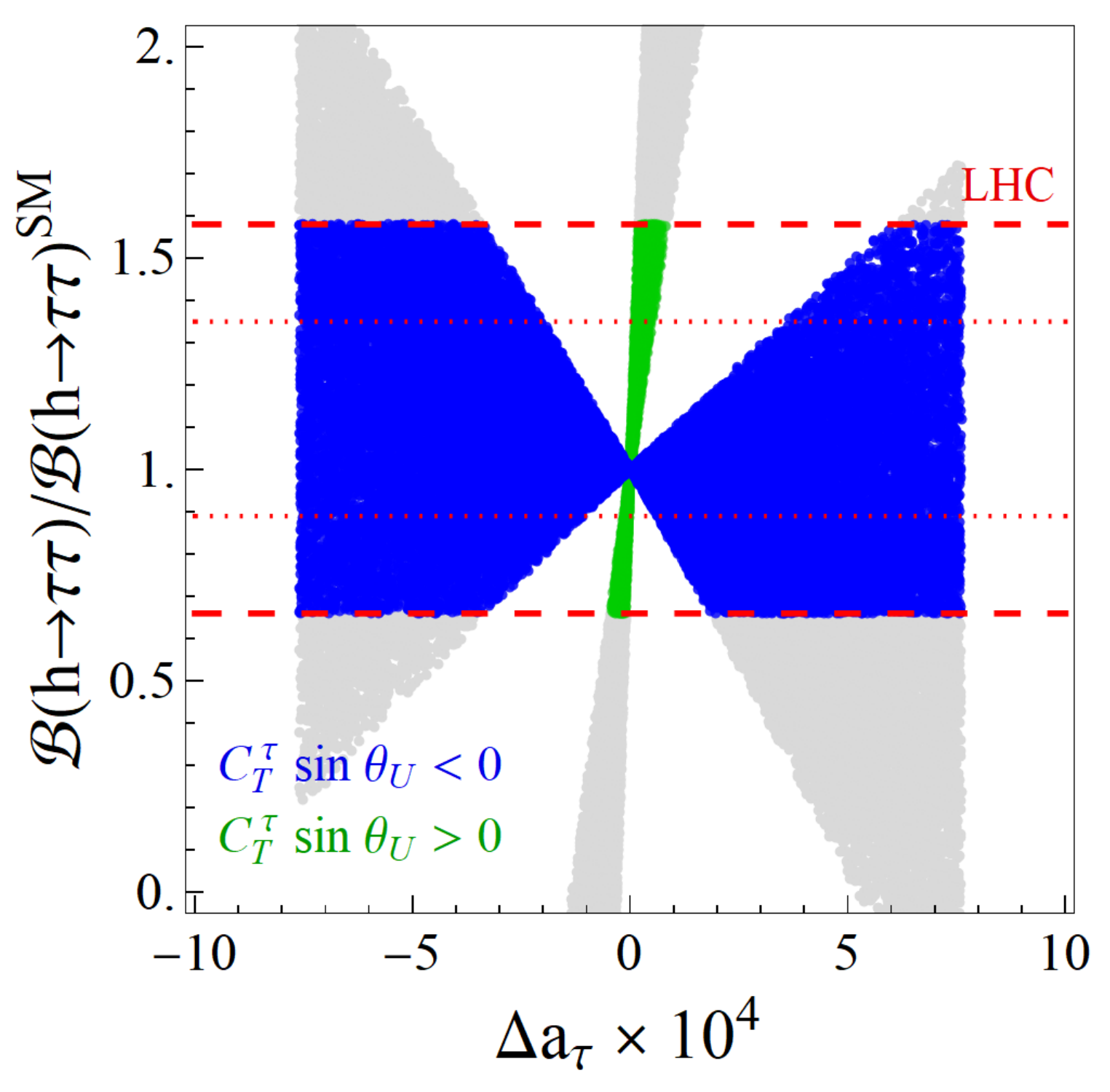}
\caption{\small \sl $\Delta a_\tau$ is plotted against $\mathcal{B}(h\to\tau\tau)/\mathcal{B}(h\to\tau\tau)^{\mathrm{SM}}$ for the points allowed by $(g-2)_\mu$ and $R_{D^{(\ast)}}$ to $1\,\sigma$ accuracy. Color code is the same as in Fig.~\ref{fig:results-couplings}, i.e.~blue and green points are allowed by all observables, including $(g-2)_\mu$ and $R_{D^{(\ast)}}$, while the gray points are excluded by the LHC constraints on $h\to\tau\tau$, c.f.~Eq.~\eqref{eq:htautau-exp}.}
\label{fig:results-correlation}
\end{figure}

The results of our numerical scan are shown in Fig.~\ref{fig:results-couplings}, where we plot the allowed values of $\sin \theta_U$ against $C_{S_L}^\tau$ (left panel) and $C_T^\tau$ (right panel). The disconnected solutions in Fig.~\ref{fig:RD-RDst-1TeV} are distinguished by different colors: (i) green for $C_T^\tau \,\sin \theta_U >0$, and (ii) blue  for $C_T^\tau  \,\sin \theta_U< 0$. LHC constraints on $\mathcal{B}(h\to \tau\tau)$ are also superimposed on the same plot, setting a lower limit on $|\sin \theta_U|$, as shown by the red dotted ($1\,\sigma$) and dashed ($2\,\sigma$) lines, c.f.~Eq.~\eqref{eq:htautau-exp}. Remarkably, this experimental result, which is still not very precise, is already useful to constraint our parameter space.  The allowed parameters in Fig.~\ref{fig:results-couplings} are then used to predict the correlation of $(g-2)_\tau$ with $\mathcal{B}(h\to\tau\tau)/\mathcal{B}(h\to\tau\tau)^{\mathrm{SM}}$ in Fig.~\ref{fig:results-correlation}, which has different features for each solution of $R_{D^{(\ast)}}$ in Fig.~\ref{fig:RD-RDst-1TeV}:

\begin{itemize}
\item[(i)] The  solution with $C_T^\tau \,\sin\theta_U>0$, shown in green, predicts a sharp correlation between $\mathcal{B}(h\to \tau\tau)$ with $\Delta a_\tau$. In particular, $|\Delta a_\tau|$ can be as large as $\approx 8 \times 10^{-5}$, while $\mathcal{B}(h\to \tau\tau)$ can saturate the current experimental limit.
\item[(ii)] On the other hand, the solution with $C_T^\tau\,\sin\theta_U <0$, shown in blue, allows for  larger values of $|\Delta a_\tau|\lesssim 8\times 10^{-4}$ due to the larger value of $|C_T^\tau|$ with respect to the previous case. Furthermore, this scenario can be perfectly consistent with the SM prediction for $\mathcal{B}(h\to \tau\tau)$, while still producing a larger effect in $\Delta a_\tau$, as depicted in Fig.~\ref{fig:results-correlation}. Our maximal prediction for this observable is only one order of magnitude below current sensitivity, and possibly within reach of Belle-II, offering an alternative to test this scenario at low-energy experiments.
\end{itemize}

\begin{figure}[p!]
\centering
\includegraphics[width=0.48\linewidth]{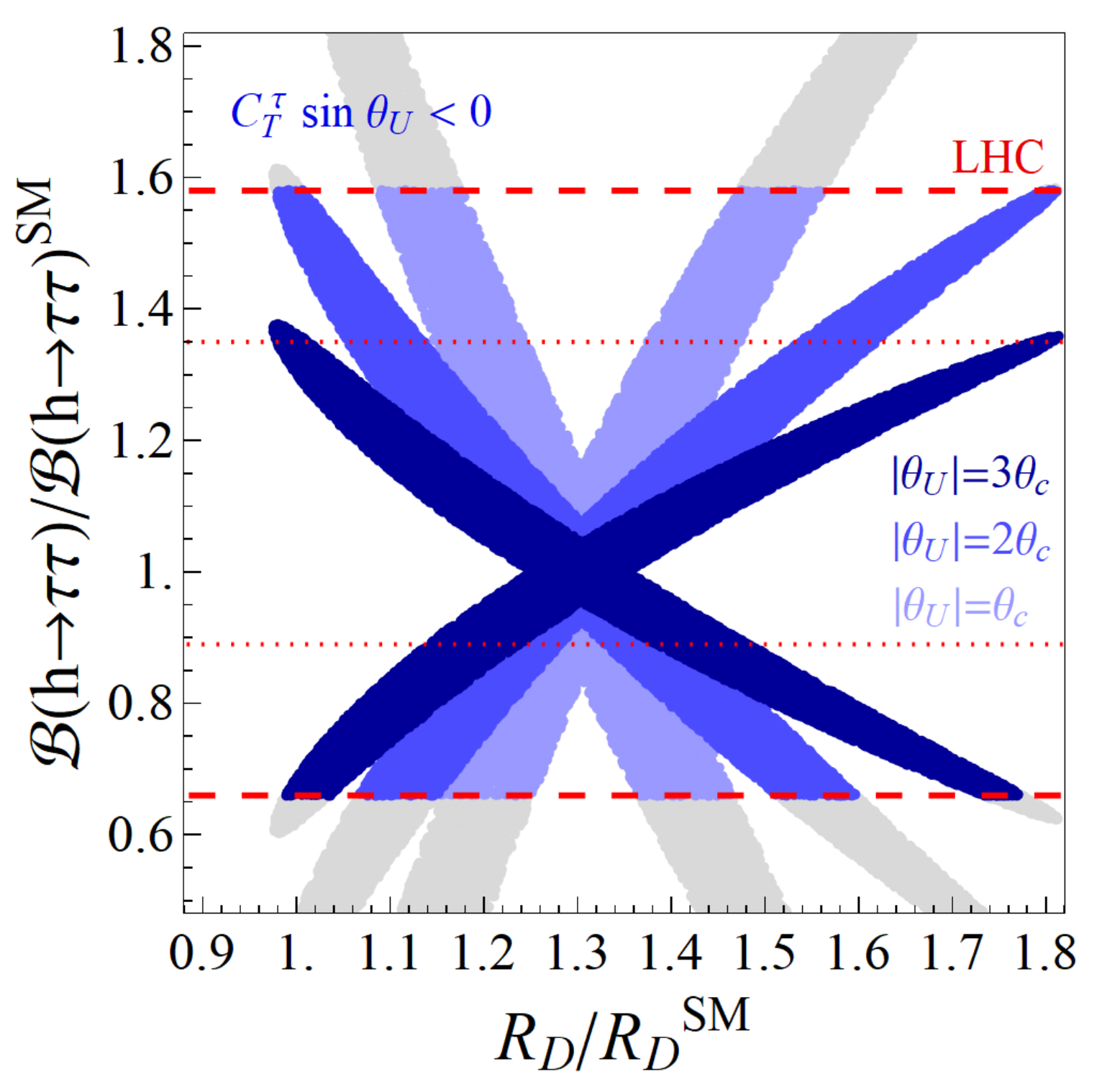}\quad\includegraphics[width=0.48\linewidth]{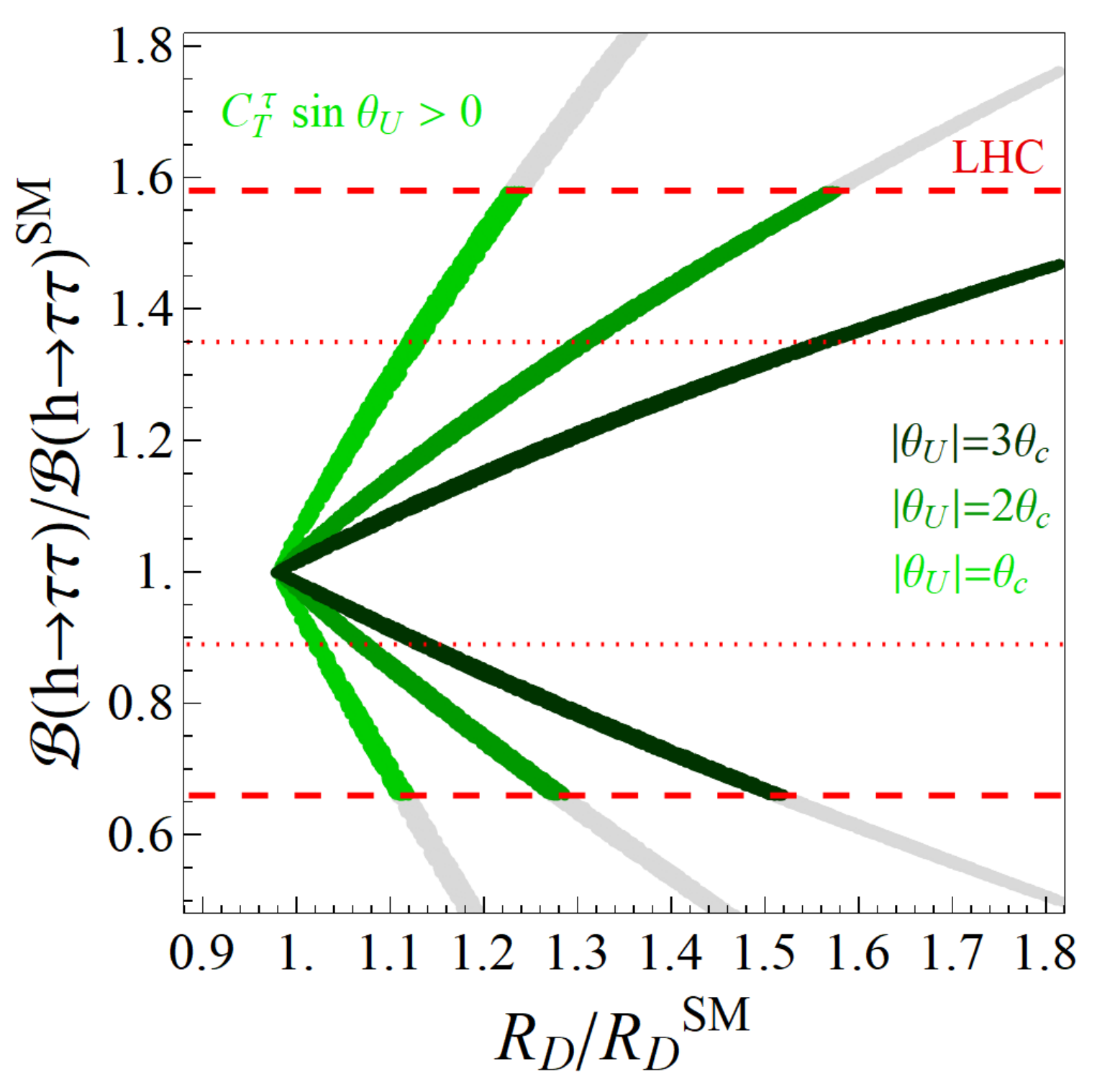}
\caption{\small \sl $R_{D}/R_{D}^{\mathrm{SM}}$ is plotted against $\mathcal{B}(h\to\tau\tau)/\mathcal{B}(h\to\tau\tau)^{\mathrm{SM}}$ for the parameters allowed by $R_{D^{(\ast)}}$ to $2\,\sigma$ accuracy, with fixed values $|\theta_U|\in \lbrace \theta_c, 2\theta_c, 3\theta_c\rbrace$. Gray points are excluded by the LHC results on the strength signal $\mu_{\tau\tau}$ [c.f.~Eq.~\eqref{eq:htautau-exp}], as in Fig.~\ref{fig:results-couplings}. See text for details.}
\label{fig:results-corellation-fixedangle-1}
\end{figure}

\begin{figure}[p!]
\centering
\includegraphics[width=0.49\linewidth]{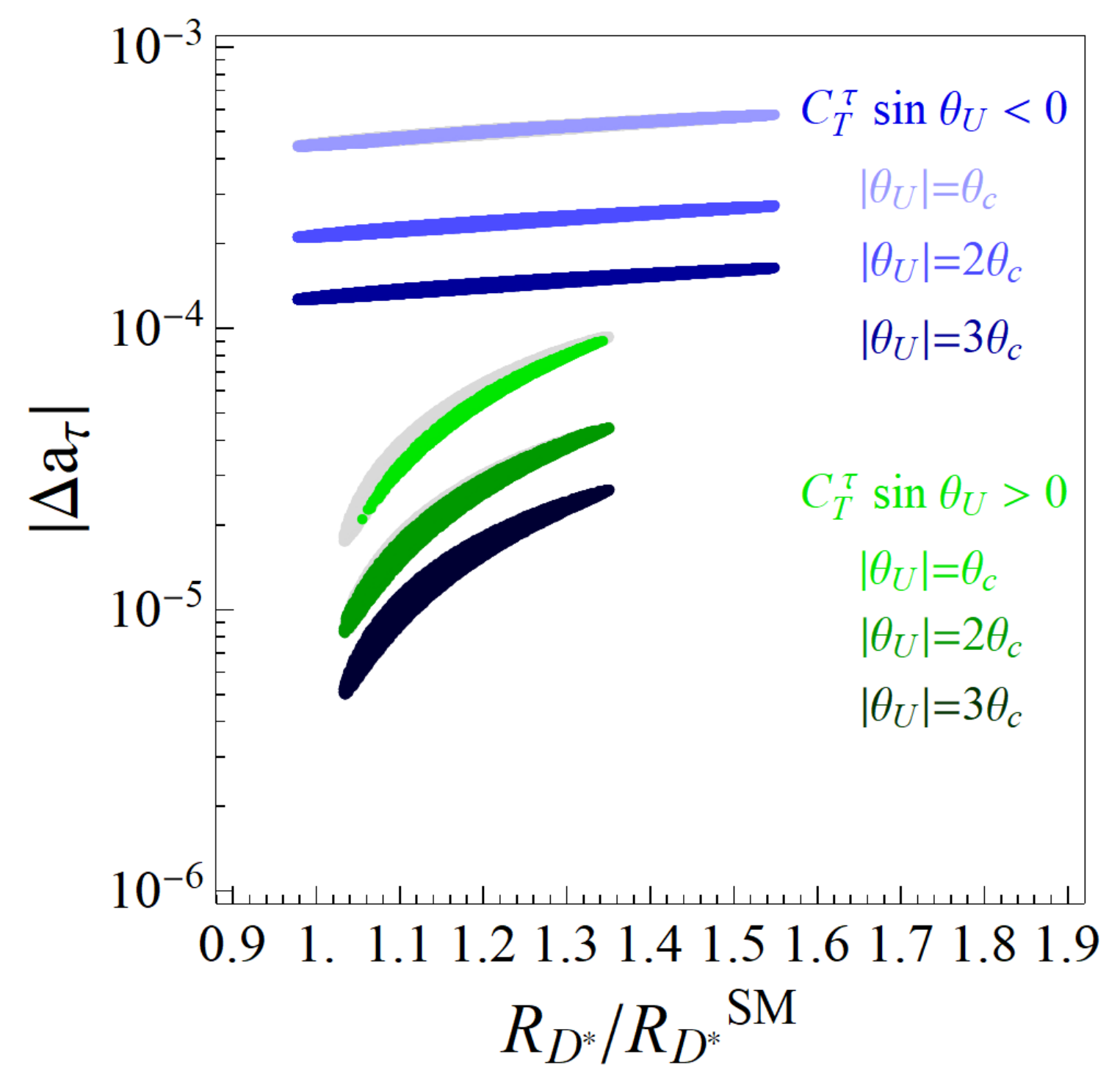}
\caption{\small \sl $R_{D^\ast}/R_{D^\ast}^{\mathrm{SM}}$ is plotted against $|\Delta a_\tau|$ for the parameters allowed by $R_{D^{(\ast)}}$ to $2\,\sigma$ accuracy, with fixed values $|\theta_U|\in \lbrace \theta_c, 2\theta_c, 3\theta_c\rbrace$. Color code is the same as in Fig.~\ref{fig:results-corellation-fixedangle-1}. }
\label{fig:results-corellation-fixedangle-2}
\end{figure}

Before concluding this Section, we stress that our findings depend on our flavor assumptions, and, in particular, on the angle $\theta_U$. To illustrate the $\theta_U$ dependence of our predictions we perform a scan of parameters similar to the one described in Eq.~\eqref{eq:scan}, but with fixed values $|\theta_U|\in\lbrace\theta_c,\,2\theta_c,3\theta_c\rbrace$, where $\theta_c$ is the Cabibbo angle $\theta_c\approx 13^\circ$. The parameters consistent with our fit of $R_{D^{(\ast)}}$ to $2\,\sigma$ accuracy are then used to correlate (i) $R_{D}/R_{D}^{\mathrm{SM}}$ with $\mathcal{B}(h\to \tau\tau)/\mathcal{B}(h\to \tau\tau)^{\mathrm{SM}}$, and (ii) $R_{D^\ast}/R_{D^\ast}^{\mathrm{SM}}$ with $\Delta a_\tau$, in Figs.~\ref{fig:results-corellation-fixedangle-1} and \ref{fig:results-corellation-fixedangle-2}, respectively. From these plots, we learn that the correlation among these observables is different for the two solutions of $R_{D^{(\ast)}}$ found in Fig.~\ref{fig:RD-RDst-1TeV}, which are depicted by the blue and green points. Furthermore, we note that the induced loop effects are more pronounced for small values of $\theta_U$, as expected from Eq.~\eqref{eq:dip-bis} and \eqref{eq:yuk-eff}, since $\lambda^{uu}_{33}=\cos \theta_U$. This relation also implies that the large top-quark enhancement of these observables can be avoided if $\theta_U \approx \pi/2$. In other words, our predictions rely on the assumption that the effective coefficients are hierarchical in flavor space, with the $b\to c\tau \bar{\nu}$ contribution being generated via the mixing angle $\theta_U$.

\section{Specific New Physics models}
\label{sec:NP_models}

The discussion above was based on the minimal assumption that only the operators in Eq.~\eqref{eq:LNP0} arise at the scale $\Lambda \approx 1~\mathrm{TeV}$. In this Section we discuss to which extent our conclusions are respected by concrete NP scenarios. In Table~\ref{tab:states}, we list the field content that can generate the effective operators in Lagrangian~\eqref{eq:LNP0} after tree-level matching to the SM effective field theory~\cite{Buchmuller:1985jz,Grzadkowski:2010es,deBlas:2017xtg}. These particles are classified in terms of their spin and SM quantum numbers. In the same table, we also collect for each model the other semileptonic four-fermion operators arising from the matching. From this table, we learn that:

\begin{itemize}
   \item[•] Extensions of the SM Higgs sector such as two-Higgs doublet models can only generate the operators $\mathcal{O}_{\ell e qd}$ and/or $\mathcal{O}_{\ell e q u}^{(1)}$, which cannot provide a viable explanation of $R_{D^{(\ast)}}$ due to the $B_c$-meson lifetime constraint discussed in Sec.~\ref{ssec:RD}, c.f.~also~Ref.~\cite{Celis:2016azn}.
   \item[•] Vector LQ bosons can induce the scalar operator $\mathcal{O}_{\ell e qd}$, as well as $\mathcal{O}_{\ell q}^{(1)}=\big{(}\bar{L} \gamma_\mu L\big{)} \big{(}\bar{Q} \gamma^\mu Q\big{)}$ and $\mathcal{O}_{\ell q}^{(3)}=\big{(}\bar{L} \gamma_\mu \tau^I L\big{)} \big{(}\bar{Q} \gamma^\mu \tau^I Q\big{)}$. The operators $\mathcal{O}_{\ell e q u}^{(1)}$ and $\mathcal{O}_{\ell e q u}^{(3)}$, characterized by the enhancement of chirality suppressed observables via up-type quark contributions, are not present in these models.
   \item[•] The only particles capable of generating specific combinations of $\mathcal{O}_{\ell e q u}^{(1)}$ and $\mathcal{O}_{\ell e q u}^{(3)}$ are scalar LQs with quantum numbers $R_2=(\mathbf{3},\mathbf{2},7/6)$ and  $S_1=(\mathbf{\bar{3}},\mathbf{1},1/3)$, which we will now discuss in detail.
\end{itemize}

\noindent Furthermore, we emphasize that tensor operators are not a
peculiarity of LQ models. They can also arise from radiative corrections and non-perturbative 
effects.

\begin{table}[htbp!]
\renewcommand{\arraystretch}{1.6}
\centering
\begin{tabular}{|c|c|c|ccc:cccc|}\hline
Field  & Spin & Quantum Numbers &   $C_{\ell e qd}$ & $C_{\ell e q u}^{(1)}$ & $C_{\ell e q u}^{(3)}$ & $C_{\ell u}$ & $C_{qe}$ & $C_{\ell q}^{(1)}$ & $C_{\ell q}^{(3)}$\\ \hline\hline
$H^\prime$  & $0$ & $(\mathbf{1},\mathbf{2},1/2)$ &  $\checkmark$ & $\checkmark$ & -- & -- & -- & -- & -- \\ 
$S_1$  & $0$ & $(\mathbf{\bar{3}},\mathbf{1},1/3)$ & --  & $\checkmark$ & $\checkmark$ & -- & -- & $\checkmark$ & $\checkmark$\\ 
$R_2$  & $0$ & $(\mathbf{3},\mathbf{2},7/6)$ &  -- & $\checkmark$ & $\checkmark$ & $\checkmark$ & $\checkmark$ & -- & --\\ \hline
$U_1$ & $1$ & $(\mathbf{3},\mathbf{1},2/3)$ & $\checkmark$ & -- & -- & -- & -- & $\checkmark$ & $\checkmark$\\ 
$V_2$  & $1$ & $(\mathbf{\bar{3}},\mathbf{2},5/6)$ &  $\checkmark$ & -- & -- & -- & $\checkmark$ & -- & --\\ \hline
\end{tabular}
\caption{ \sl \small Classification of the particle fields that can generate the operators $\mathcal{O}_{\ell e qd}$, $\mathcal{O}_{\ell e qu}^{(1)}$ and $\mathcal{O}_{\ell e qu}^{(3)}$ by tree-level matching, in terms of the SM quantum numbers, $(SU(3)_c,SU(2)_L,Y)$, with $Q=Y+T_3$ as hypercharge convention. For completeness, we also list the other semileptonic operators generated by these scenarios in the Warsaw basis~\cite{Buchmuller:1985jz,Grzadkowski:2010es}.}
\label{tab:states} 
\end{table}

The scalar LQ $R_2$ was proposed as a viable candidate to explain $R_{D^{(\ast)}}$ in Ref.~\cite{Sakaki:2013bfa,Becirevic:2018uab,Becirevic:2018afm}. The possibility of using the same state to accommodate the muon $g-2$ and $R_{K^{(\ast)}}$ anomalies has also been explored in Ref.~\cite{ColuccioLeskow:2016dox} and~\cite{Becirevic:2017jtw}, respectively. Remarkably, $R_2$ is the only scalar LQ that automatically preserves baryon number~\cite{Assad:2017iib}. The most general Yukawa Lagrangian allowed in this scenario by gauge symmetry reads~\cite{Dorsner:2016wpm}

\begin{align}
\label{eq:yukawa-R2}
\mathcal{L}_{R_2} = y_R^{ij} \, \overline{Q^a}_i e_{R\,j}\,R_2^a + y_L^{ij} \, \overline{u}_{R\,i}  L_j^a \, \varepsilon_{ab}\,{R_2}^b + \mathrm{h.c.}\,,
\end{align}

\noindent where $y_L$ and $y_R$ are generic couplings, $i,j$ are flavor indices, and $a,b$ are $SU(2)_L$ indices. After integrating out the scalar LQ, we obtain at the matching scale
\begin{align}
\label{eq:leff-R2}
\begin{split}
\mathcal{L}_{\mathrm{eff}}^{R_2} =&-\dfrac{y_R^{ij}\big{(}y_R^{kl}\big{)}^\ast}{2\,m_{R_2}^2} \big{[}\mathcal{O}_{qe}\big{]}_{iklj}-\dfrac{y_L^{ij}\big{(}y_L^{kl}\big{)}^\ast}{2\,m_{R_2}^2} \big{[}\mathcal{O}_{\ell u}\big{]}_{ljik} \\
&-\Bigg{[}\dfrac{y_R^{ij}\big{(}y_L^{kl}\big{)}^\ast}{2\, m_{R_2}^2}\Bigg{(}\big{[}\mathcal{O}_{\ell equ}^{(1)}\big{]}_{ljik}+ \dfrac{1}{4}\big{[}\mathcal{O}_{\ell equ}^{(3)}\big{]}_{ljik}\Bigg{)}+\mathrm{h.c.}\Bigg{]}\,,
\end{split}
\end{align}
\noindent where $m_{R_2}$ is the LQ mass, and $\mathcal{O}_{qe}=\big{(}\bar{Q} \gamma_\mu Q\big{)} \big{(}\bar{e}_R \gamma^\mu e_R\big{)}$ and $\mathcal{O}_{\ell u}=\big{(}\bar{L} \gamma_\mu L\big{)} \big{(}\bar{u}_R \gamma^\mu u_R\big{)}$ belong to the Warsaw basis~\cite{Buchmuller:1985jz,Grzadkowski:2010es}. This equation tells us that $C_{\ell e q u}^{(1)}=4\, C_{\ell e q u}^{(3)}$ at $\mu = m_{R_2}$, which can be translated into $g_{S_L} \approx 8.12 \, g_T$ at $\mu=m_b$. This particular combination of couplings lies outside the $2\,\sigma$ region in Fig.~\ref{fig:RD-RDst-1TeV} and cannot account for $R_{D^{(\ast)}}^{\mathrm{exp}}>R_{D^{(\ast)}}^{\mathrm{SM}}$ with real couplings.~\footnote{The possibility of explaining $R_{D^{(\ast)}}$ with complex Yukawa couplings of the LQ state $R_2$ has been previously considered in Ref.~\cite{Sakaki:2013bfa,Becirevic:2018uab,Becirevic:2018afm}, in which case this LQ state becomes a viable candidate.} Note that to generate these effects, it is unavoidable to induce nonzero contributions to $C_{\ell u}$ and $C_{qe}$, which will then contribute to $Z$-pole observables via loop effects, as already discussed in Ref.~\cite{Becirevic:2018uab}. Nonetheless, it is worth emphasizing that the potential signature in $(g-2)_\tau$ and the sizable modification of the $h\to \tau\tau$ decay are a novelty of our work, which were not considered in previous LQ studies.

Lastly, the scalar LQ $S_1$ was considered as a candidate to accommodate $R_{D^{(\ast)}}$ in Ref.~\cite{Sakaki:2013bfa,Bauer:2015knc,Becirevic:2016oho,Cai:2017wry,Marzocca:2018wcf}. The most general Yukawa Lagrangian in this case reads~\cite{Dorsner:2016wpm}

\begin{align}
\label{eq:yukawa-S1}
\begin{split}
\mathcal{L}_{S_1} &= x_L^{ij} \, \overline{Q^{C\,a}} L_j^b\,\varepsilon_{ab}\, S_1 + x_R^{ij} \,\overline{u^C_{R\,i}} e_{R\,j}\, S_1\\[0.4em]
&+ z_L^{ij} \,\overline{Q^{C\,a}_{i}} Q_{L\,j}^b\,\epsilon_{ab}\, S_1^\ast + z_R^{ij} \,\overline{u^C_{R\,i}} d_{R\,j}\, S_1^\ast +\mathrm{h.c.}\,,
\end{split}
\end{align}

\noindent where $x_{L,R}$ and $z_{L,R}$ are generic Yukawa matrices, and the superscript $C$ stands for the charge conjugation. Differently from the previous scenario, this model can induce the proton decay via the dangerous diquark couplings $z_L$ and $z_R$, which must therefore be avoided, for example by enforcing a suitable symmetry. By neglecting the diquark couplings we obtain at the matching scale
\begin{align}
\label{eq:leff-S1}
\begin{split}
\mathcal{L}_{\mathrm{eff}}^{S_1} =&\dfrac{x_R^{ij}\big{(}x_R^{kl}\big{)}^\ast}{2\,m_{S_1}^2} \big{[}\mathcal{O}_{eu}\big{]}_{ljki}+\dfrac{x_L^{ij}\big{(}x_L^{kl}\big{)}^\ast}{4\,m_{S_1}^2}\Bigg{(} \big{[}\mathcal{O}_{\ell q}^{(1)}\big{]}_{ljki}-\big{[}\mathcal{O}_{\ell q}^{(3)}\big{]}_{ljki}\Bigg{)} \\
&+\Bigg{[} \dfrac{\big{(}x_L^{ij}\big{)}^\ast x_R^{kl}}{2 m_{S_1}^2}\Bigg{(}\big{[}\mathcal{O}_{\ell equ}^{(1)}\big{]}_{jlik}- \dfrac{1}{4}\big{[}\mathcal{O}_{\ell equ}^{(3)}\big{]}_{jlik}\Bigg{)}+\mathrm{h.c.}\Bigg{]}\,,
\end{split}\end{align}

\noindent where $m_{S_1}$ is the LQ mass, and $\mathcal{O}_{e u}=\big{(}\bar{e}_R \gamma_\mu e_R\big{)} \big{(}\bar{u}_R \gamma^\mu u_R\big{)}$. Interestingly, this scenario generates $C_{\ell e q u}^{(1)}=-4\, C_{\ell e q u}^{(3)}$ and $C_{\ell q}^{(1)}=-C_{\ell q}^{(3)}$, which correspond to $g_{S_L}=-4 \, g_T$ and $g_{V_L}$ in Eq.~\eqref{leff:bc}. Both of these combinations can independently accommodate $R_{D^{(\ast)}}$. Similarly to the previous model, the operators $\mathcal{O}_{\ell q}^{(1)}$ and $\mathcal{O}_{\ell q}^{(3)}$ can induce additional effects both at tree and loop-level which have already been extensively studied in Ref.~\cite{Feruglio:2016gvd,Feruglio:2017rjo,Cornella:2018tfd}. Note, in particular, that a scenario with large scalar and tensor operators can be obtained from this model if $|x_L^{ij}|\ll |x_R^{ij}|$, c.f.~Ref.~\cite{Sakaki:2013bfa,Cai:2017wry}. 

While in the framework of the EFT analyzed in the previous sections a simultaneous explanation of $R_{D^{(\ast)}}$ and the muon $g-2$ is made possible 
by a specific choice of the Wilson coefficients, this conclusion cannot be reached without  additional investigations in the context of scalar LQ models. Indeed, from LQ exchange, the pure third generation coupling to quarks and leptons assumed throughout this note and the coupling to the third quark generation and the second lepton generation, 
necessary to explain the muon $g-2$, necessarily imply potentially dangerous couplings 
involving both the second and the third lepton generations. The latter can lead to a rate of 
$\tau\to \mu\gamma$ above the current limit, unless a fine-tuning of the involved parameters 
is made.

\section{Conclusion}
\label{sec:conclusion}
Recent data on charged-current semileptonic decays $B\to D^{(*)} \ell \nu$ hint at a lepton flavor universality (LFU) violation departing from the Standard Model (SM) predictions at the 4\,$\sigma$ level. If confirmed, these anomalies will represent a major New Physics (NP) discovery. Since the required amount of LFU violation is quite large, correlated signals are expected to emerge also in other low- and/or high-energy observables, depending on the features of the underlying NP scenario. An UV complete model supplying an explanation to the charged-current anomaly
would clearly represent the ideal self-consistent framework to analyse all possible correlated signals. While many interesting proposals exist in the literature,
present data are not yet sufficient to single out a unique scenario and a considerable room is left for analyses based on an effective field theory (EFT) approach,
which can capture the main features of the underlying theory. In particular, if the NP scale $\Lambda$ is of order TeV, the appropriate formalism
is that of an EFT invariant under the SM gauge group, selecting a very limited number of relevant semileptonic operators, up to flavor combinations.
An EFT setup of this type is particularly suitable to study the unavoidable IR effects generated, through RGE flow, at low energy.
Due to operator mixing the low-energy theory encompasses entirely new classes of effects related to the semileptonic operators,
independently on whether or not they are present at the UV level.

In this work, we focused on a NP setup defined at the scale $\Lambda \gg v$ by scalar and tensor semileptonic operators, which have 
been proven to be able to accommodate the charged-current anomalies. We have adopted a conservative mixing pattern in the quark sector,
by assuming NP coupling to the third generation and the minimum amount of mixing needed to explain $R_{D^{(\ast)}}$. 
In the lepton sector we have considered a slightly more general pattern, with couplings to all generations, initially taken diagonal.
In a first step, we have derived the RGE of our effective Lagrangian from the high-scale 
$\Lambda$ down to lower energies along the lines of Ref.~\cite{Feruglio:2016gvd,Feruglio:2017rjo,Cornella:2018tfd}. 
Then, we have outlined the phenomenological implications of our setup to be confirmed or disproved by future data. 
In particular we have identified in the parameter space of the model a viable region allowing a solution to the anomalies
in terms of a weighted combination of the scalar and tensor operators $\mathcal{O}^{(1)}_{\ell e qu}$ and $\mathcal{O}^{(3)}_{\ell e qu}$. 

 We found that the scalar operator $\mathcal{O}^{(1)}_{\ell e qu}$ generates modifications of the Higgs couplings to leptons via RGE-induced 
electroweak effects. Despite the loop suppression, these effects can compete with the SM contributions due to the large chiral enhancement $m_t/m_\ell$. Experimentally, the most promising channel to be monitored at the LHC is $h\to\tau\tau$.
The tensor operator $\mathcal{O}^{(3)}_{\ell e qu}$ generates leptonic dipole moments which are again chirally enhanced by $m_t/m_\ell$. 
While the predictions for the electron and muon $g-2$ are not directly correlated with $R_{D^{(*)}}$ since they depend 
on the leptonic couplings with light generations, the $(g-2)_{\tau}$ turns out to be of order $\mathcal{O}(10^{-3})$ in the 
parameter space where $R_{D}$ and $R_{D^*}$ are accounted for. It is worth emphasizing that this correlation depends on the flavor assumptions we have made, which maximize the effects in the top-quarks loops while inducing contributions to $b\to c\tau\bar{\nu}$ via quark mixing. In principle, one can reduce these effects by choosing a peculiar flavor structure in the quark sector. For instance, large loop effects do not arise when only electroweak singlet up-quarks of the second generation enter the scalar and tensor operators studied here. Another important outcome of our analysis concerns the pattern of NP couplings to leptons. Non-diagonal couplings and/or intergenerational mixing 
should be very small, to avoid the stringent bounds from charged lepton radiative decays. On the other hand, a sizable mixing
affecting left-handed leptons is naturally expected, given the mixing pattern observed in neutrino oscillations.
Reconciling these aspects could represent a serious challenge for flavor models.
Finally, scalar and tensor operators arise in models via tree-level LQ exchange and, for completeness, we have recalled some
concrete realizations in the last section of this work.  

Our analysis is complementary to those of Refs.~\cite{Feruglio:2016gvd,Feruglio:2017rjo,Cornella:2018tfd,Gonzalez-Alonso:2017iyc}.
Indeed, in Ref.~\cite{Gonzalez-Alonso:2017iyc}, it has been pointed out that scalar and tensor operators mix through electroweak running effects.
However, the unavoidable generation of NP contributions to leptonic magnetic moments and Higgs leptonic couplings, which is the main focus of 
the present analysis, has not been addressed.
Moreover, Ref.~\cite{Feruglio:2016gvd,Feruglio:2017rjo,Cornella:2018tfd} focused on (current)$\times$(current) operators. In that case, 
electroweak running effects showed up predominantly in $Z$-pole observables and leptonic $\tau$ decays, which are instead not significantly
modified in our setup.

In conclusion, we have shown that the scenario considered here exhibits several distinctive phenomenological features that can be tested in ongoing and upcoming experiments such as LHC and Belle-II. Since such signatures emerge only at the quantum level, we emphasize the importance of including electroweak corrections 
in any framework where the explanation of $B$-anomalies invokes NP at the TeV scale.

\section{Acknowledgments}
\label{sec:acknowledgments}

We would like to thank Massimo Passera, Svjetlana Fajfer and Kin Mimouni for useful discusssions. This project has received support in part by the MIUR-PRIN project 2015P5SBHT 003 ``Search for the Fundamental Laws and Constituents'' and by the European Union's Horizon 2020 research and innovation programme under the Marie Sklodowska-Curie grant agreement N$^\circ$~674896 and 690575. The research of P.~P.~was supported in part by the ERC Advanced Grant No. 267985 (DaMeSyFla), by the research grant TAsP, and by the INFN.

\appendix


\end{document}